\documentclass{ieeeaccess}
\usepackage{cite}
\usepackage{amsmath,amssymb,amsfonts}
\usepackage{algorithmic}
\usepackage{graphicx}
\usepackage{textcomp}
\usepackage{comment}
\usepackage{url}
\usepackage{here}
\usepackage{enumerate}

\def\BibTeX{{\rm B\kern-.05em{\sc i\kern-.025em b}\kern-.08em
    T\kern-.1667em\lower.7ex\hbox{E}\kern-.125emX}}
\begin{document}
\history{Date of publication xxxx 00, 0000, date of current version xxxx 00, 0000.}
\doi{10.1109/ACCESS.2023.0322000}

\title{ProgrammableGrass: A Shape-Changing Artificial Grass Display Adapted for Dynamic and Interactive Display Features}
\author{Kojiro Tanaka \authorrefmark{1}, Akito Mizuno \authorrefmark{1}, Toranosuke Kato \authorrefmark{1}, Masahiko Mikawa \authorrefmark{2}, Makoto Fujisawa \authorrefmark{2}}

\address[1]{Graduate School of Comprehensive Human Sciences, Tsukuba, Ibaraki, Japan}
\address[2]{Institute of Library, Information and Media Science, Tsukuba, Ibaraki, Japan}

\tfootnote{This work was supported by JST SPRING, Grant Number JPMJSP2124.}

\markboth
{Author \headeretal: Preparation of Papers for IEEE TRANSACTIONS and JOURNALS}
{Author \headeretal: Preparation of Papers for IEEE TRANSACTIONS and JOURNALS}

\corresp{Corresponding author: Kojiro Tanaka (e-mail: tanaka.kojiro.sp@alumni.tsukuba.ac.jp).}

\begin{abstract}
    There are various proposals for employing grass materials as a green landscape-friendly display. However, it is difficult for current techniques to display smooth animations using 8-bit images and to adjust display resolution, similar to conventional displays. We present \textit{ProgrammableGrass}, an artificial grass display with scalable resolution, capable of swiftly controlling grass color at 8-bit levels.  This grass display can control grass colors linearly at the 8-bit level, similar to an LCD display, and can also display not only 8-bit-based images but also videos.  This display enables pixel-by-pixel color transitions from yellow to green using fixed-length yellow and adjustable-length green grass. We designed a grass module that can be connected to other modules. Utilizing a proportional derivative control, the grass colors are manipulated to display animations at approximately 10 [fps]. Since the relationship between grass lengths and colors is nonlinear, we developed a calibration system for \textit{ProgrammableGrass}. We revealed that this calibration system allows \textit{ProgrammableGrass} to linearly control grass colors at 8-bit levels  through experiments under multiple conditions. Lastly, we demonstrate \textit{ProgrammableGrass} to show smooth animations with 8-bit  grayscale  images.  Moreover, we show several application examples to illustrate the potential of \textit{ProgrammableGrass}.   With the advancement of this technology, users will be able to treat grass as a green-based interactive display device. 
\end{abstract}


\titlepgskip=-21pt

\maketitle

\section{INTRODUCTION}

\begin{figure*}[h]
  \centering
  \includegraphics[width=\linewidth]{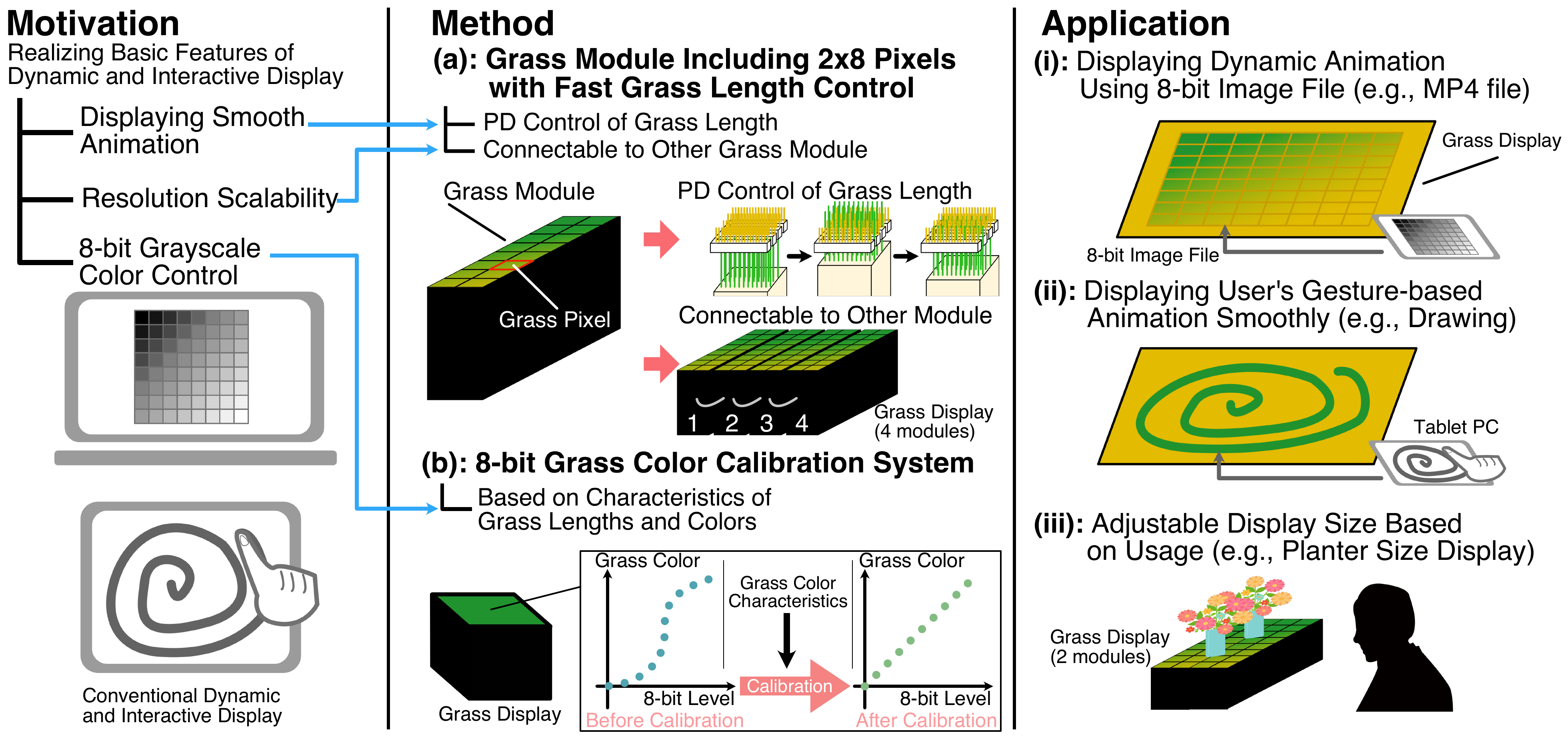}
  \caption{ Overview of (a, b) methods and (i, ii, iii) applications in \textit{ProgrammableGrass}} 
  \label{paperConcept}
\end{figure*}

In recent years, there has been an increase in services that enhance the convenience and entertainment value of green spaces, both indoors and outdoors, by displaying various images and animations using liquid crystal displays (LCDs) \cite{digitalSinage} and projectors \cite{mosimosi}. However, while these technologies offer exciting possibilities, they also present challenges. For example, LCDs have an impact on the natural beauty of green space due to their artificial appearance \cite{gavin}. Projectors, on the other hand, are difficult to use in the bright conditions typically found in these spaces. 

As a result, there is growing interest in image display methods that can maintain green space while displaying images in bright environments, such as those using grass materials. For example, using a grass mowing method, an illustration of Beethoven was created that blended well with green space \cite{grassmow}. Traditionally, grass-based image display methods often relied on primitive approaches, such as directly mowing the grass. Recently, these methods have been developed to print images on a grass surface by adjusting the grass shading or applying paint to grass \cite{sugiura_grassffiti_2017,new_ground_technology_airprint_2022,ackroyd__harvey_big_2007}. Furthermore,  we previously have proposed several grass display techniques to play animations \cite{tanaka_natural_2021,tanaka_dynamic_2023, mizuno_2023}. In the grass techniques, the colors of artificial grass can be changed by moving grass lengths. However, current techniques are still unable to display 8-bit-based images and animations or provide a high response speed. Contemporary displays, including LCDs and projectors, mainly use 8-bit levels to dynamically control colors, matching the 8-bit levels in common image and video file formats like PNG and MP4. However, this standard is not directly applicable to conventional grass display systems due to controlling grass color in only several levels.  In addition, current grass display hardware systems can hardly adjust their resolution. When these problems can be solved, grass materials can be handled more like a dynamic and interactive display similar to an LCD or a projector. 

We present \textit{ProgrammableGrass}, a shape-changing artificial grass display with the resolution scalability to swiftly control the grass color at 8-bit levels. In this grass display, the color of the grass can be dynamically controlled at 8-bit levels, similar to an LCD, enabling the display of 8-bit-based image and video materials.  Figure \ref{paperConcept}(a) and (b) show the overview of the methods in \textit{ProgrammableGrass}. In the grass display, the grass color changes from yellow to green pixel by pixel by moving the green grass vertically among the yellow grass. The pixel system is called a grass pixel. The grass pixel controls the grass length using direct current (DC) motors equipped with rotary encoders. It also includes the fast Proportional-Derivative (PD) control of the DC motor.    As shown in Figure \ref{paperConcept}(a), to adjust resolution of \textit{ProgrammableGrass}, we design a grass module with $2\times8$  grass  pixels that can be connected to other grass modules. In this study, we connect four grass modules to create \textit{ProgrammableGrass} of $8\times8$ pixels. 

Furthermore, as the relationship between grass lengths and colors is nonlinear and varies slightly from pixel to pixel, we design an 8-bit grass color calibration system  as shown in Figure \ref{paperConcept}(b). The color of the grass pixel is captured at fixed grass length intervals using a digital camera. Then, after image processing, the color change relative to the grass length is represented by CIEDE2000, an equation that incorporates a human visual model to denote color differences \cite{sharma_ciede2000_2005}. Based on this grass color change relative to the grass length, a correspondence table between the 8-bit levels and the grass lengths is determined, ensuring that the change in the grass color relative to the 8-bit levels is linear. By applying this correspondence table to the grass pixel, the grass color can be controlled at 8-bit levels. Moreover, by considering the characteristics of multiple grass pixels simultaneously, it is possible to control the color of the grass linearly while minimizing color variation among multiple grass pixels.
 We develop a tablet tool for the calibration system to facilitate the grass calibration process. We conduct indoor experiments to test whether the calibration system can bring 8-bit linear grass color control to \textit{ProgrammableGrass} while also minimizing color variation across grass pixels. 

As shown in Figure \ref{paperConcept}(i), (ii), and (iii), we construct several animation and application examples to demonstrate the potential of \textit{ProgrammableGrass}.  In the demonstrations, \textit{ProgrammableGrass} can play dynamic animation using 8-bit grayscale MP4 video files, and show gesture-based animations via a table computer.

This paper's contributions of include:

\begin{itemize}
  \item Design and development of the grass module hardware of \textit{ProgrammableGrass} for fast PD grass length control and resolution scalability.
  \vskip.5\baselineskip
  \item Design of the 8-bit grass color calibration system based on the relationship between grass lengths and colors across multiple grass pixels, and development the tablet tool for the calibration system.
  \vskip.5\baselineskip
  \item Evaluation of the 8-bit grass color calibration system in  several lighting conditions including a standard color evaluation environment.
  \vskip.5\baselineskip
  \item Demonstrations of animations and applications of \textit{ProgrammableGrass} to show its performance and potential.
\end{itemize}

\section{RELATED WORK}

\subsection{Non-Luminescent Displays}

In addition to grass materials, there are many display methods that use the unique colors of natural materials. Display methods using liquids have been proposed, focusing on water-wet ground or asphalt \cite{nagafuchi_polka_2020,designboom_nicholas_2011}, bubbles and colors created by electrolysis \cite{ishii_bubbowl_2019, ishii_electrolysis_2020}, condensation \cite{tsujimoto_ketsuro-graffiti_2016}, liquid-metal droplets \cite{sahoo_tangible_2018}, and tubes filled with water \cite{inoue_tuve_2018}. Since carpet shading can be easily changed by simply rubbing one's finger on a carpet, display methods using carpets have also been proposed \cite{sugiura_graffiti_2014,yamamoto_robotic_2022}. In addition, Yamamoto et al. presented a carpet method to switch images according to the positions of a viewer or a light source \cite{yamamoto_turning_2023}. Photo-Chromeleon is a method that changes multi-color textures using photochromic dyes, and the color of the materials can be switched with the light of a certain wavelength \cite{jin_photo-chromeleon_2019}. Organic display methods focusing on moss \cite{kimura_moss-xels_2014}, mimosa \cite{gentile_plantxel_2018}, and foodstuffs \cite{robinson_sustainabot_2019} have also been published. In the field of art, Daniel Rozin has created multiple artworks focusing on interactive display devices using natural materials such as wood and homespun baskets \cite{daniel_rozin_wooden_1999,daniel_rozin_weave_2007}. In addition, the people of Inakadate village in Japan create rice paddy art using differences between the colors of rice paddies \cite{tohoku_tourism_promotion_organization_inakadate_2022}.

The advantage of these display methods using natural materials is that they are compatible with our daily living environments.  In this paper, we focus on display methods of grass materials and propose a novel green space-friendly display technique that shows dynamic and interactive animations using grass color control.

\subsection{Pin-Array Displays}

Pin-array display techniques, in which multiple pins are controlled synchronously, have been developed primarily in the research field of shape-reproducing displays \cite{arakawa_bulkscreen_2020,follmer_inform_2013}. However, pin-array displays are also used for other purposes. For example, Nakagaki et al. proposed a hands-on animated craft platform using a pin-array display to move crafts such as dolls \cite{nakagaki_animastage_2017}. Suzuki et al. presented a dynamic and instant three-dimensional (3D) printing method that uses a pin-array display \cite{suzuki_dynablock_2018}. Haptic devices with a pin-array display have been developed, focusing on palm size \cite{nakagaki_inforce_2019,abler_hedgehog_2021} and fingertip size \cite{luo_development_2001}. Furthermore, pin-array displays are used as virtual reality (VR) tools \cite{yoshida_pocopo_2020,siu_shapeshift_2018,gonzalez_x-rings_2021,je_elevate_2021}. Nakagaki et al. represented a platform that can convert a pin-array display into other interfaces, such as a haptic display and a 3D modeling tool. Engert et al. proposed a platform for shape-changing spatial displays using a pin-array display \cite{engert_straide_2022}.

These studies benefit from a pin-array display's ability to control a shape on a point-by-point basis. In this paper, we present a method that uses a pin-array display system, including DC motors, to move the grass length pixel by pixel, allowing for quick control of the grass color.

\subsection{Grass Displays}

As with other non-luminescent display methods, grass-based display techniques have been developed in research, commerce, and art. Examples of grass display methods are pressing or mowing the grass. Nvidia generated an image of an electronic component on a grass-like green field using plants pressed against it \cite{nvidia_impossibly_2014}. Scheible et al. proposed a large-scale grass art method that used a drone to get a bird's-eye view of the grass while mowing it \cite{scheible_dronelandart_2016}. Other techniques involve using grass shading, depending on the orientation of the grass tips. Sugiura et al. presented a grass printer device that uses multiple rods to apply pressure to grass and change the direction of the grass tips. The change in direction of the grass tips enables the production of a binary image on the grass \cite{sugiura_grassffiti_2017}. Using a similar technique, NEW GROUND TECHNOLOGY performs a grass print service that uses air pressure to change the direction of the grass tips to show a large-scale image \cite{new_ground_technology_airprint_2022}.  This method uses a spatial dither technique to represent the shading in the input image by adjusting the placement of light and dark grass shades. In this method, a large number of pixel shades are required on a grass surface.  Another grass display method uses grass growth rate differences to generate a rich image. Ackroyd \& Harvey created artworks to show photography using grass \cite{ackroyd__harvey_big_2007}. Canvas containing grass seeds are exposed to the light of a negative image of photography from a projector. As the grass grows, the photography is generated with differences in grass growth rates. These methods show static images on the grass, and a shape-changing grass system is needed to switch images quickly and play animations.  We previously developed a grass display that can dynamically change the grass color by adjusting the grass length using two colors of grass \cite{tanaka_natural_2021,tanaka_dynamic_2023}.  However, the current grass display cannot swiftly control the grass color linearly at 8-bit levels, and its resolution cannot be adjustable.  

\begin{figure}[h]
  \includegraphics[width=\linewidth]{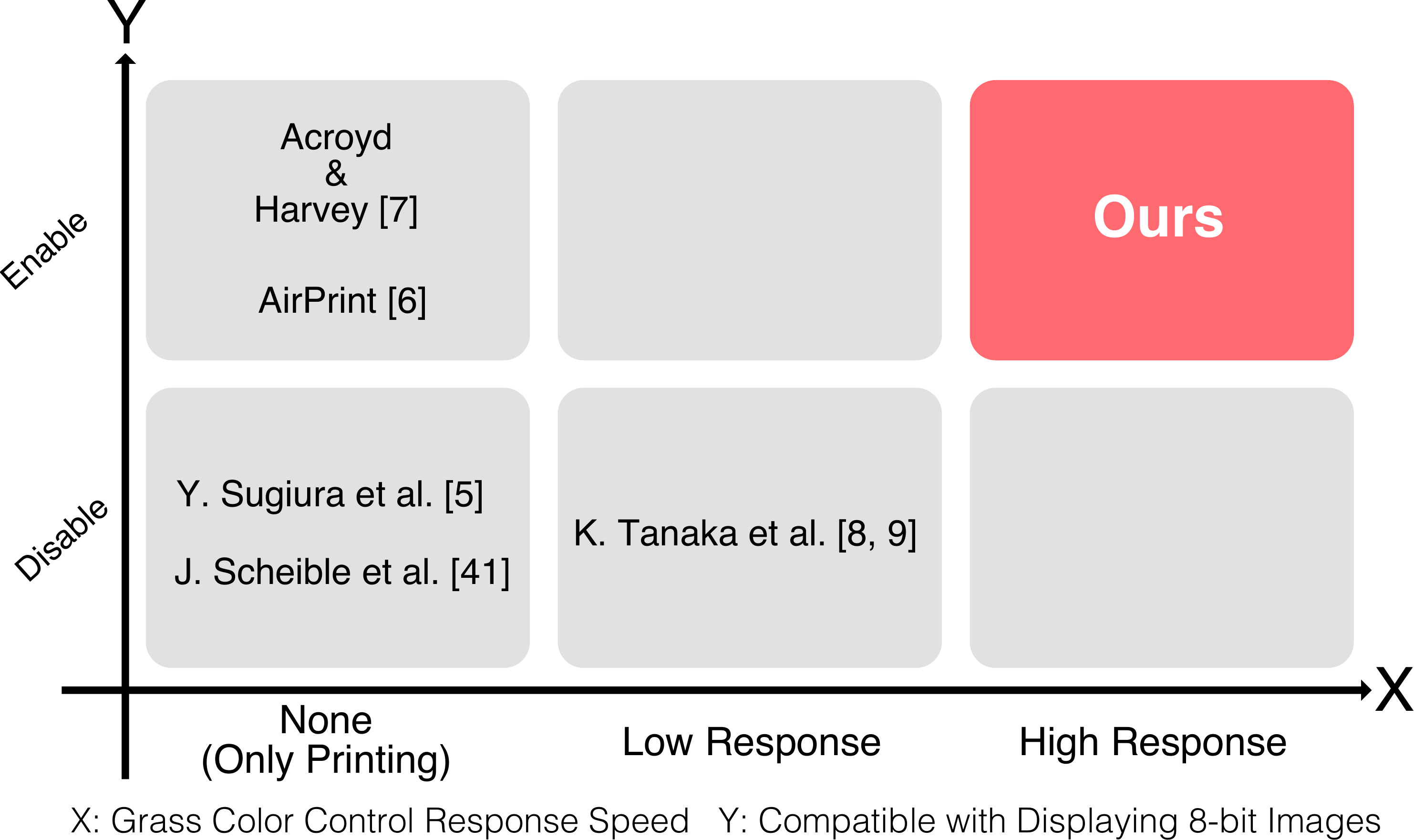}
  \caption{Position of our method relative to previous grass studies} 
  \label{position}
\end{figure}

In this paper, we propose a novel approach that quickly controls the grass color at 8-bit levels with a high response speed. These technologies enable an artificial grass display to play dynamic animations using 8-bit grayscale images such as an MP4 file, and gesture-based animations from a user. We also design a hardware module for the grass display that can be connected to other modules for easy adjustment of the resolution.  Figure \ref{position} shows the position of our work relative to previous grass studies.  Table \ref{comparison} shows the comparison between our previous work and the current work, focusing on levels of grass color control, frame rate, number of pixels, and resolution scalability. 

\subsection{Plant Color Evaluation with CIEDE2000 and 8-bit Color Control}

Plant color evaluation methods are important to develop grass color control. One of the algorithms used to evaluate plant color is CIEDE2000, a color difference formula published by the International Commission on Illumination (CIE, Commission Internationale de l'Éclairage) \cite{sharma_ciede2000_2005}. The CIEDE2000 is suitable for plant color evaluations because the formula reflects the perceptual model of human eyes. For example, leaf color evaluation systems with the CIEDE2000 have been represented to estimate nitrogen levels \cite{haider_computer-vision-based_2021,tao_smartphone-based_2020}, while an apple color evaluation method was proposed to measure the discoloration of apple pulp using the CIEDE2000 \cite{shimizu_high-throughput_2021}. A dyed wood color evaluation method also uses the CIEDE2000 \cite{wu_color_2021}. 
 
 We previously proposed a CIEDE2000-based grass display setting system to map several levels to the grass lengths for controlling the green-yellow grass color \cite{tanaka_dynamic_2023, mizuno_2023}.  This method sets levels according to whether the grass colors can be distinguished by the human eye. However, in the method, only several levels can be set, and the number of levels varies depending on differences in the environment in which the grass display is installed and the grass color itself. In other words, the current method does not allow the grass display to be calibrated to show the 8-bit images.

On the other hand, in conventional displays such as LCDs, gamma correction is a method of linearly controlling color at 8-bit levels. Gamma correction aims to linearize the relationship between the input 8-bit levels and luminance, taking into account the non-linear characteristics between the display's luminance and voltage. This color calibration technique, originally developed for Cathode Ray Tube (CRT) displays \cite{CRT}, has been adapted and explored for use with other types of displays as well. This includes research on methods for LCDs \cite{LCD1, LCD2, LCD3, LCD4} and Organic Light-Emitting Diode (OLED) displays \cite{OLED1}. These studies have been dedicated to developing calibration methods specifically tailored to the unique characteristics of each type of display. Moreover, the widespread adoption of standardized color management systems, such as International Color Consortium (ICC) profiles \cite{ICC_profiles}, has significantly facilitated the consistent management of gamma correction across various displays.

 However, these calibration schemes are primarily designed based on the luminance characteristics of self-emitting displays. Our proposed grass display is one of reflective displays, including electronic papers. Reflective displays are significantly influenced by external environmental conditions in color control \cite{reflective}. Furthermore, the measurement methods for characteristics of grass color in relation to the grass length are not well established. Consequently, for controlling grass color at 8-bit levels, a calibration approach based on an external environment and a characteristic between grass colors and lengths is important.

 We present a novel CIEDE2000-based stable calibration system that allows for linear control of grass color at 8-bit levels using the characteristic of grass color variation with grass lengths. The calibration system also takes into account differences between the colors of multiple grass pixels. Furthermore, we develop a calibration tool using a tablet computer to easily facilitate the calibration of \textit{ProgrammableGrass}.

\begin{table}[h]
  \centering
  \caption{Comparison with previous grass animation display system}
  \begin{tabular}{|c|c|c|}
  \hline
  \multicolumn{1}{|l|}{} & Previous Work & Ours     \\ \hline
  Level of Grass Color Control                  & Several       & 8 bit (256)    \\ \hline
  Frame Rate             & 1 fps         & 10 fps   \\ \hline
  Numbers of Pixel       & 9             & 64 \\ \hline
  Resolution Scalability & Unable        & Able     \\ \hline
  \end{tabular}
  \label{comparison}
\end{table}

\vspace{10pt}

\section{DESIGN AND IMPLEMENTATION OF GRASS MODULE}
\label{HardModule}
\subsection{Principle}

\begin{figure}[h]
  \includegraphics[width=\linewidth]{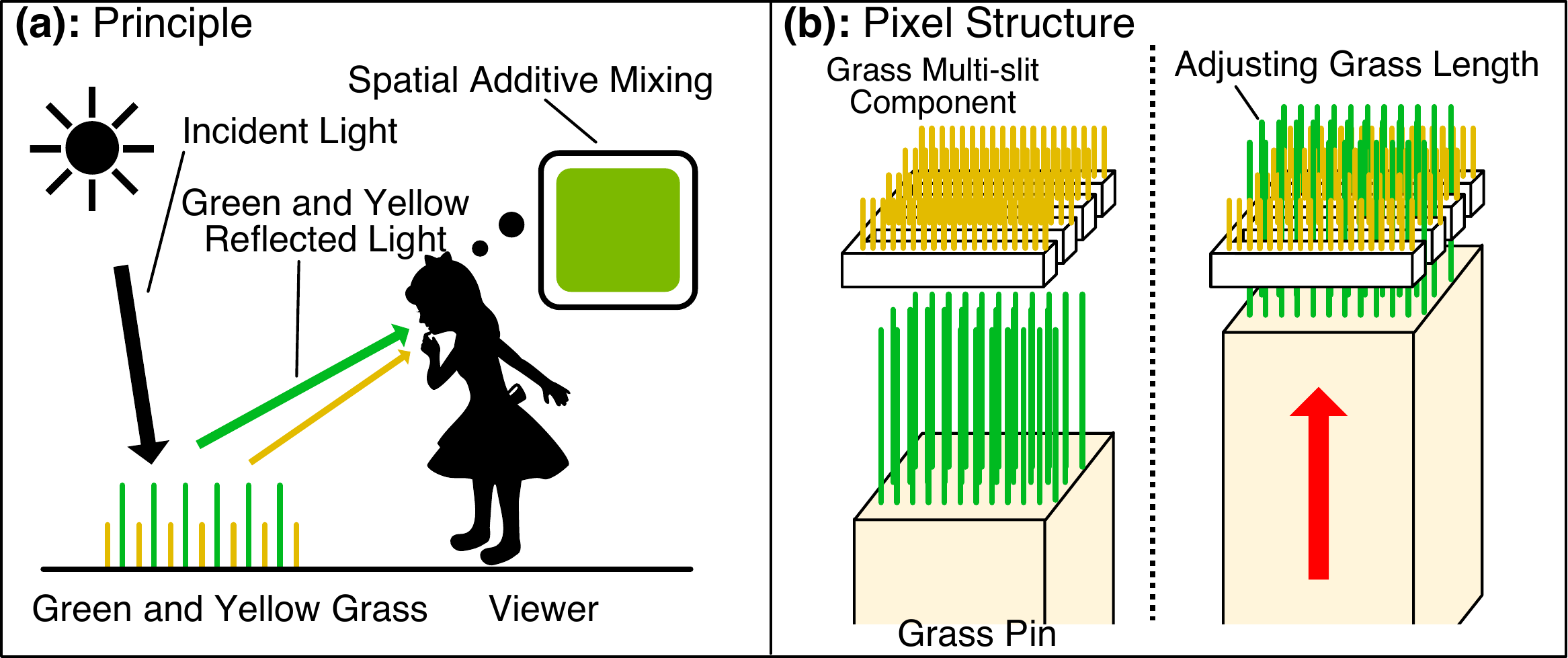}
  \caption{(a) Principle of spatial additive mixing in grass color and (b) Pixel structure using grass pin and grass multi-slit component} 
  \label{principle}
\end{figure}

The grass color of \textit{ProgrammableGrass} is determined by the mixture of green and yellow grass, utilizing the principle of spatial additive mixing as shown in Figure \ref{principle}(a). Spatial additive mixing is a phenomenon in which humans perceive a texture's color as the average color when the texture is too fine for the human eyes to distinguish. This phenomenon often occurs when humans see grass surfaces, including lively green grass and dying yellow grass. 

 Using this phenomenon,  we previously developed the pixel structure called a grass pixel \cite{tanaka_natural_2021,tanaka_dynamic_2023} as shown in Figure \ref{principle}(b). The pixel structure comprises a green grass pin and a yellow grass multi-slit component. The green grass is attached to the top surface of a pin, and the yellow grass is planted on the top surface of a multi-slit component. By passing the green grass through the gaps of the yellow grass and moving the green grass pins vertically, the pixel structure appears like lively grass growing out among dying grass, thereby changing the perceived grass color. In this study, we adopted the grass pixel as the pixel system for \textit{ProgrammableGrass}.

\subsection{Hardware Structure}
\label{sectionhard}

\begin{figure}[h]
  \centering
  \includegraphics[width=0.8\linewidth]{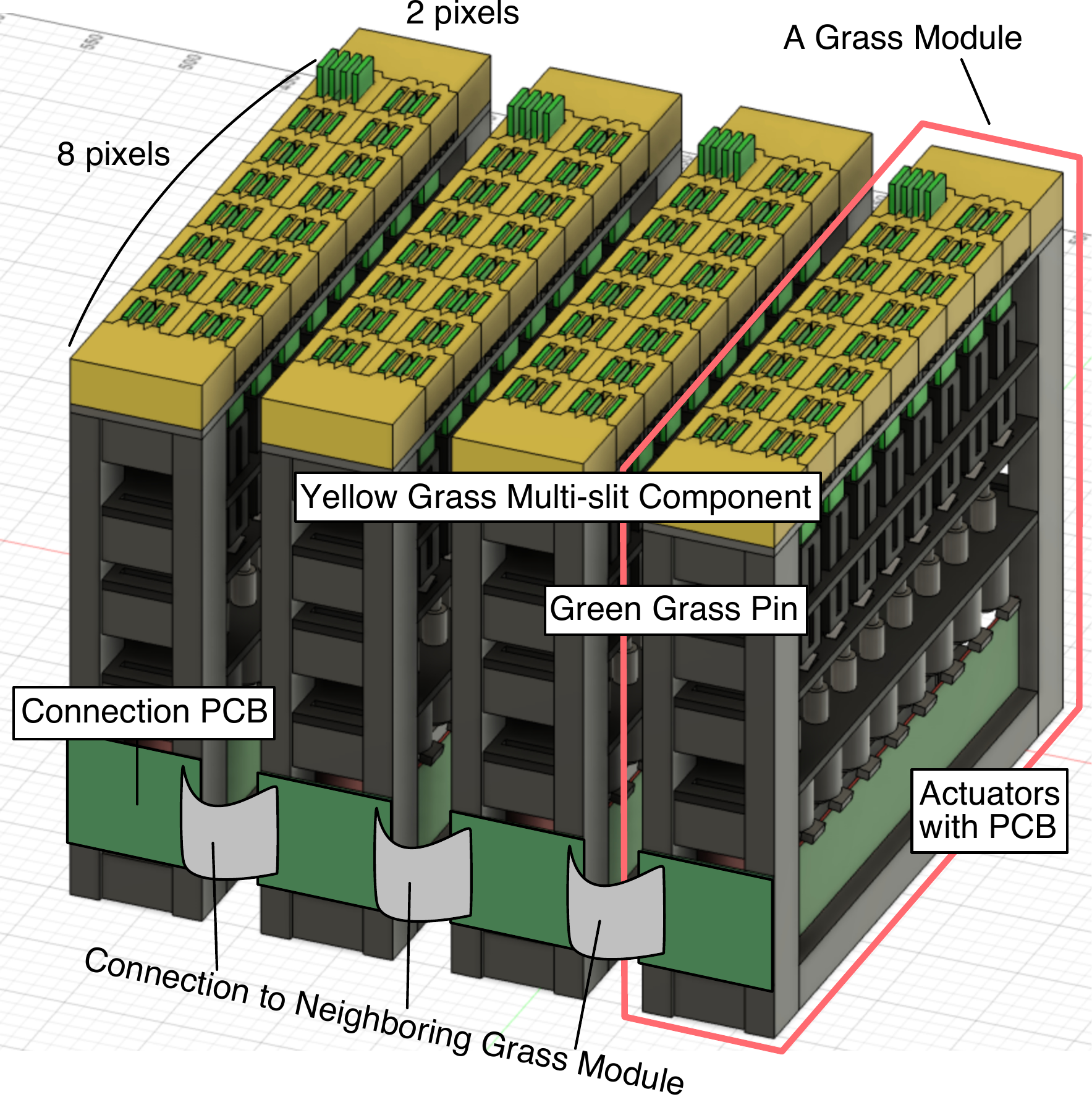}
  \caption{Hardware concept design of grass module created by 3D modeling software} 
  \label{concept}
\end{figure}

We designed a module that is the smallest unit of \textit{ProgrammableGrass} to facilitate the scalability of the resolution. The module was named a grass module. Figure \ref{concept} shows the overview of the grass module hardware concept design. The resolution of the grass module is $2\times8$ pixels, and the grass module includes green grass pins, yellow grass multi-slit components, and actuators with printed circuit boards (PCBs). In addition, the grass module has a PCB to connect with another grass module. Multiple grass modules can be chained together via the connection PCB to increase the width of the resolution of \textit{ProgrammableGrass}. For example, when two grass modules are chained, the resolution is $4\times8$ pixels and \textit{ProgrammableGrass} can be used as a small artificial grass display. When four grass modules are chained, the resolution is $8\times8$ pixels and the \textit{ProgrammableGrass} can show graphical animations such as text and icons. In this paper, we built four grass modules and assembled a \textit{ProgrammableGrass} with a resolution of $8\times8$ pixels.

\begin{figure}[h]
  \centering
  \includegraphics[width=0.8\linewidth]{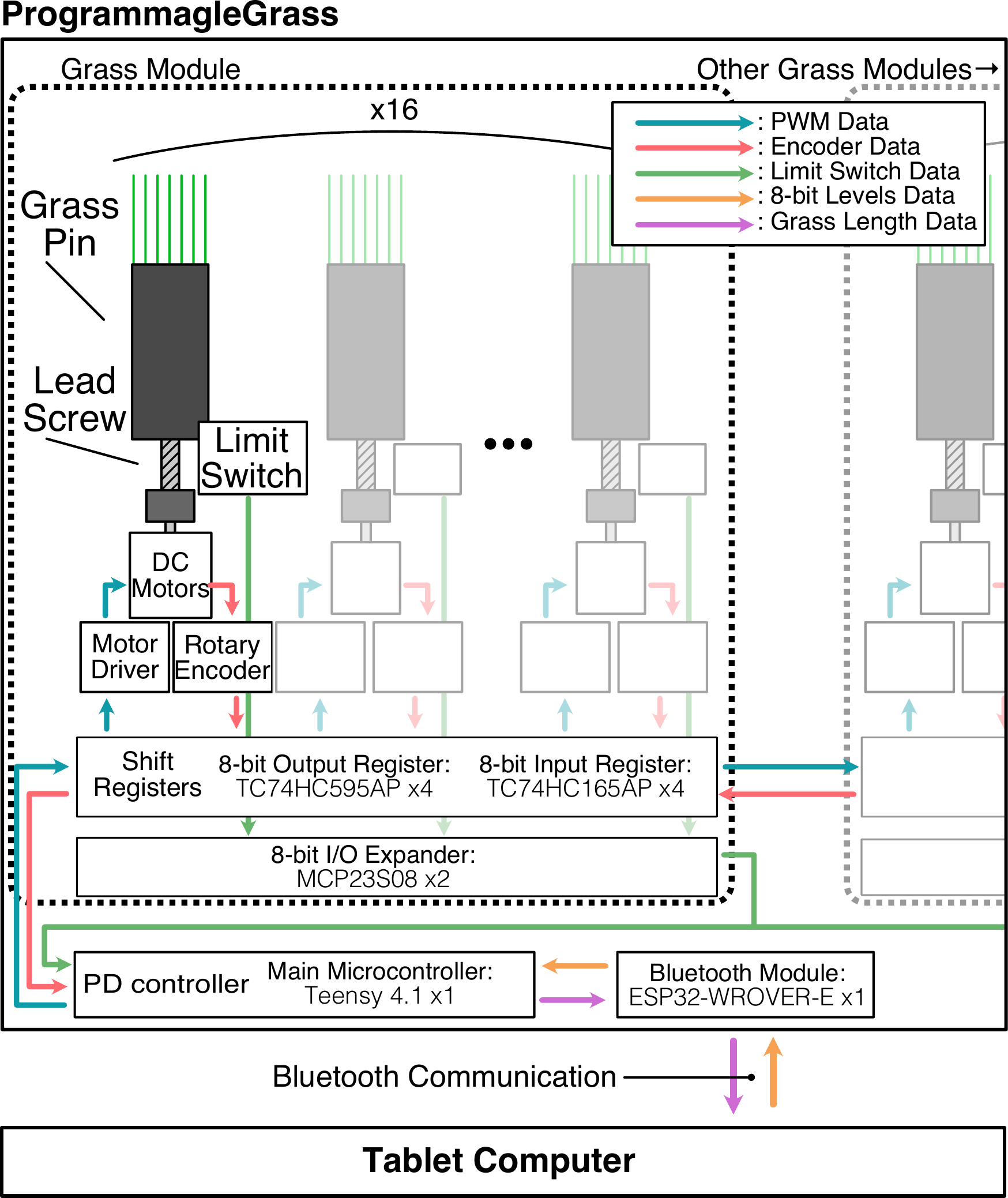}
  \caption{Hardware system overview of grass module}
  \label{sysoverview}
\end{figure}

Figure \ref{sysoverview} shows the overall hardware system of \textit{ProgrammableGrass}. The grass module has 16 grass pins, and a DC motor actuates each grass pin via a lead screw (lead: 6 [mm/rotation]). The DC motor is controlled by a full-bridge DC motor driver (TB6643KQ). A magnetic rotary encoder tracks the rotary position of the DC motor. The rotary encoder's resolution for the DC motor and the lead screw combination is 10/73 [mm/count]. A limit switch is used to check the origin position of the grass pin. The grass module is operated by a microcontroller (Teensy 4.1). The microcontroller sends pulse width modulation (PWM) values to the motor drivers using 8-bit output shift registers (TC74HC595AP). In addition, the microcontroller receives the rotary positions of the DC motors using 8-bit input shift registers (TC74HC165AP). The states of limit switches of the grass pins are sent to the microcontroller through serial peripheral interface (SPI) I/O (Input and Output) expanders (MCP23S08). These shift registers and SPI devices are included in the grass module. Since the shift registers can be cascaded with other shift registers and the I/O expanders can share SPI signal lines, the grass modules can be connected to other grass modules, and several grass modules can then be controlled by the microcontroller.  Moreover, a Bluetooth module (ESP32-WROVER-E) is connected to the microcontroller via SPI signal lines, inputting an 8-bit level and outputting the grass length for each pixel wirelessly.  For example, this module allowed \textit{ProgrammableGrass} to be operated wirelessly with a tablet computer during the calibration in Section \ref{calibTool} and the demonstration in Section \ref{demo}. The grass module uses 3.3V supplies to operate the shift registers and I/O expanders, and 12V supplies to run the DC motors. The voltages are shared among several grass modules.

\subsubsection{Grass Pixel}

\begin{figure}[h]
  \centering
  \includegraphics[width=0.7\linewidth]{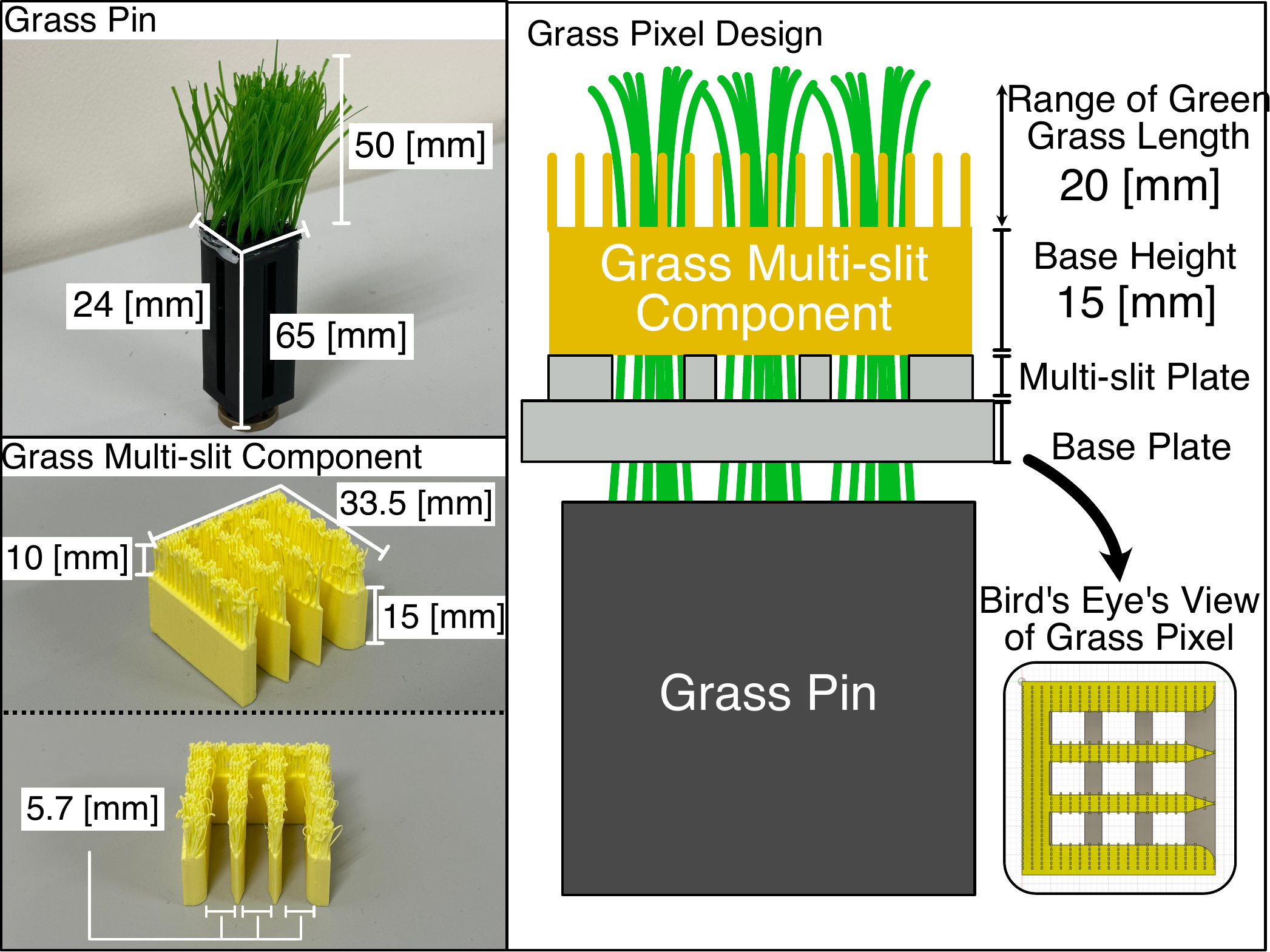}
  \caption{Grass pixel hardware design including green grass pin and yellow grass multi-slit}
  \label{Pixel}
\end{figure}

Figure \ref{Pixel} shows the grass pixel hardware design of \textit{ProgrammableGrass}. In the grass pin, a black pin is 3D printed, and its size is $24\times24\times65$ [mm]. Artificial green grass (length: 50 [mm]) is planted on the top surface of the black pin. The grass multi-slit component is 3D printed, including the yellow grass. Its surface area is $33.5\times33.5$ [mm], and the base height of the component is 15 [mm]. The length of the yellow grass is 10 [mm]. The width of each slit is 5.7 [mm]. The range of the green grass length is designed to be from 0 to 20 [mm] based on the base surface of the grass multi-slit component. The multi-slit plate (size: $33.5\times33.5\times2.0$ [mm]) allows the green grass to come out evenly through the yellow grass. The multi-slit plate has three slits (width: 5.7 [mm]) and the same slit structure as the grass multi-slit component. When the grass multi-slit component and the multi-slit plate are superimposed to be perpendicular to each other, the holes to pass the green grass become grids, as shown in the grass pixel design image of Figure \ref{Pixel}. The green grass is first inserted into the multi-slit plate and then into the grass multi-slit component. Therefore, the green grass can be evenly moved amid the yellow grass. The grass multi-slit component and the multi-slit plate are placed on a base plate (thickness: 5 [mm]) that has a square hole to pass the green grass (area: $25\times25$ [mm]).

\subsubsection{Grass Module}

\begin{figure}[h]
  \centering
  \includegraphics[width=0.7\linewidth]{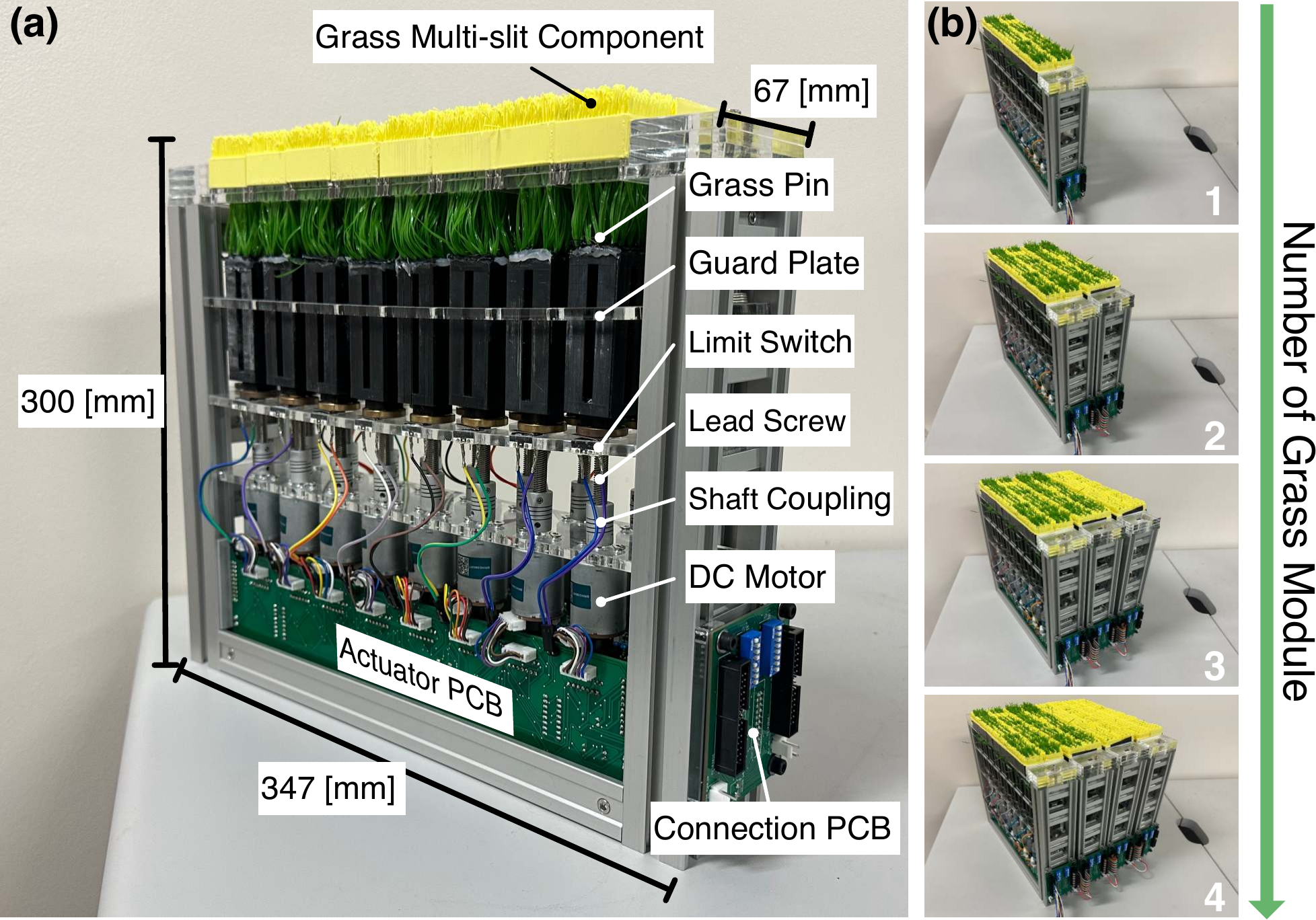}
  \caption{(a) Grass module hardware design and (b) Resolution adjustment using several grass modules}
  \label{grassmodule}
\end{figure}

Figure \ref{grassmodule}(a) shows the overview of the grass module hardware design. The size of the grass module is $300\times347\times67$ [mm]. The grass module has 16 grass pixels and contains 16 each of DC motors, shaft couplings, lead screws, limit switches, grass pins, and grass multi-slit components. A guard plate is used to move the grass pins vertically. The guard plate includes 16 square holes (area: $25\times25$ [mm]) to pass the grass pins. In the enclosure of the grass module, two actuator PCBs are placed to run the grass pins. A single actuator PCB can drive eight grass pins. A connection PCB is placed on one side of the enclosure. The connection PCB is connected with two actuator PCBs inside the enclosure and is equipped with two flat-cable sockets on the outside. The flat-cable sockets are used to connect the module with another grass module. Each of the sockets connects to the immediate neighboring grass modules. In this manner, the resolution of \textit{ProgrammableGrass} can be adjusted by combining several grass modules in Figure \ref{grassmodule}(b).

\subsection{Software Structure}
\label{sectionsoft}
\begin{figure}[h]
  \centering
  \includegraphics[width=\linewidth]{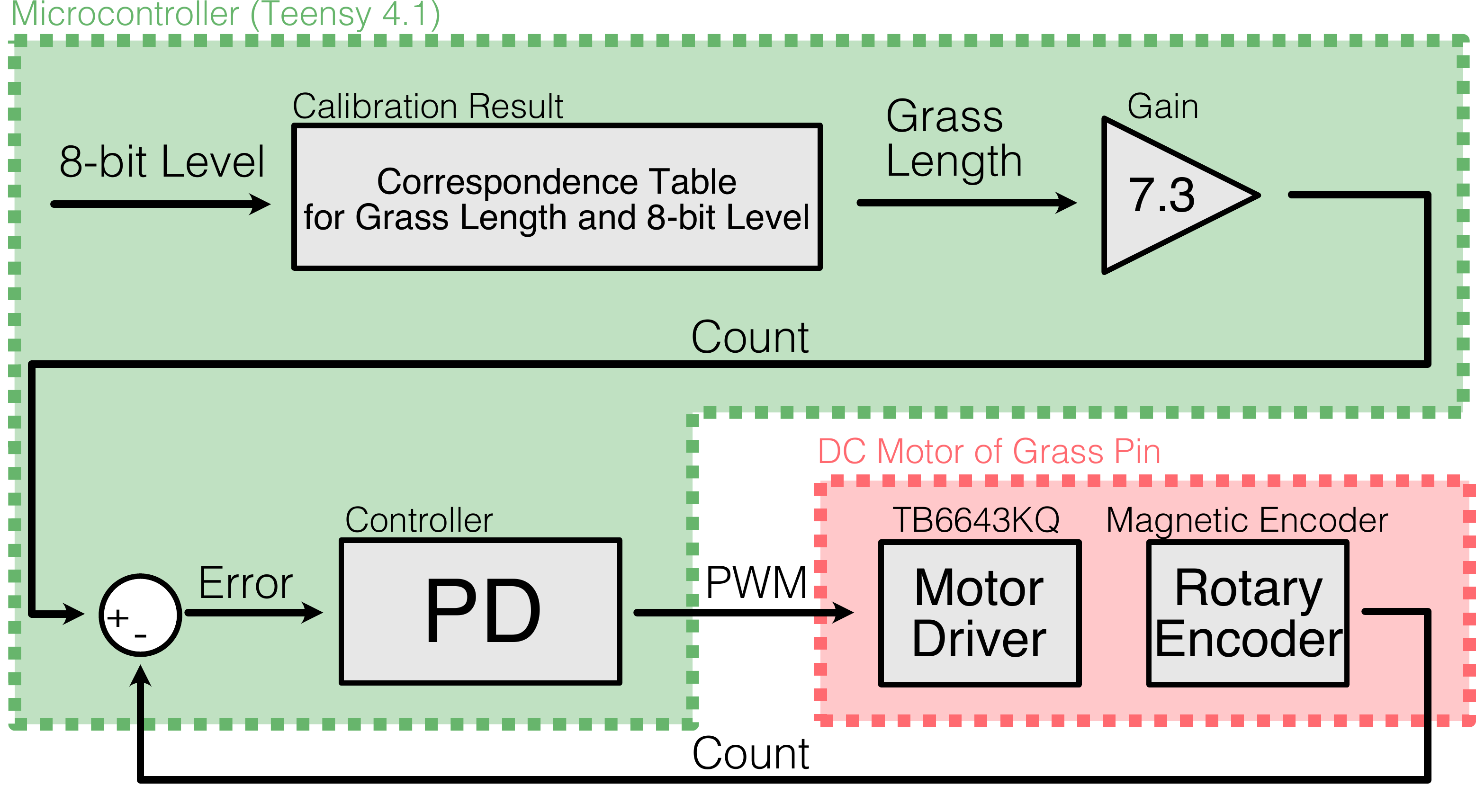}
  \caption{Software design of DC motor's PD Control with correspondence table for grass length and 8-bit level.}
  \label{pd}
\end{figure}

Figure \ref{pd} illustrates the overall software design of \textit{ProgrammableGrass}, which is developed in a C++ environment. The microcontroller controls the grass pixel using a PD controller developed with MATLAB and Simulink. The green grass length is adjusted based on the 8-bit levels. First, the 8-bit level is converted into the grass length using a correspondence table for the grass length and the 8-bit level. The correspondence table is obtained from a grass color calibration system that measures the color of the grass pixel and determines the relationship between the grass length and the 8-bit level. The details of the grass color calibration system are described in Section \ref{sectioncalib}. Second, the grass length value is converted into a count value based on the rotary encoder of the DC motor. Since the resolution of the rotary encoder is 10/73 [mm/count], the gain value is set to 7.3 [count/mm]. Then the converted count is also cast to an integer. In the PD control, the converted count value is a desired setpoint (SP), and the output count value from the rotary encoder is a process variable (PV).  In setting the PD control parameters in a heuristic manner, the P parameter is first set so that the PV oscillates near the SP, and the D parameter is set to suppress this oscillation.  The PD controller then sends a PWM value to the DC motor driver using the SP and the PV. Therefore, the grass length is adjusted based on the 8-bit levels through the PD control. This process is repeated for each grass pixel by the microcontroller.

\vspace{-20pt}

\subsection{Experiment on Response Speed}
\label{response}

\begin{figure}[h]
  \centering
  \includegraphics[width=0.7\linewidth]{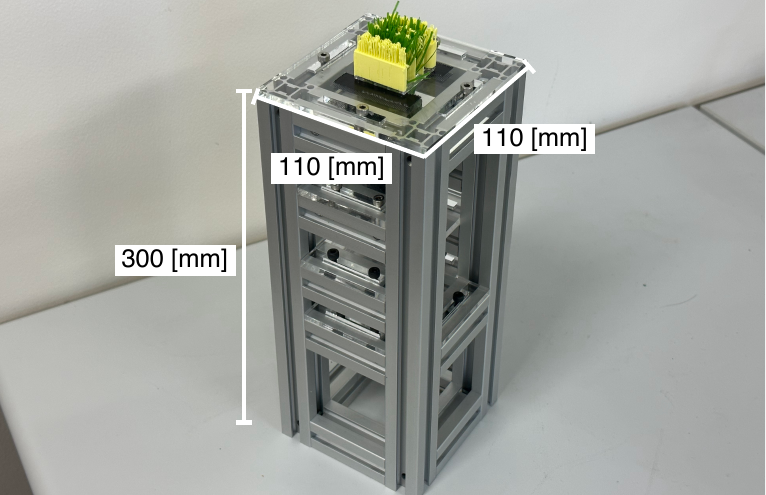}
  \caption{Grass pixel having the same system as grass module.}
  \label{singlepixel}
\end{figure}

We conducted experiments on the response speed of \textit{ProgrammableGrass}. We developed a grass pixel for the experiments as shown in Figure \ref{singlepixel}. This grass pixel was built based on the hardware and software structures for the grass module.  The P and D parameters were set at 6.0 and 10.0, respectively.  The size of the grass pixel was $300\times110\times110$ [mm]. In the experiments, we sent linear SPs to the grass pixel from 0 to  146  [count] for moving the grass length from 0 to a maximum of 20 [mm] in a fixed time interval. In addition, the PVs from the rotary encoder were measured. When the PVs follow the linear SPs, the grass pixel can control the grass length with the time interval. The time interval was set to $1/n$ [sec] corresponding to a natural number $n$ frames per second (fps). Then, the maximum response speed at which the PVs can follow the linear SPs was evaluated as the $n$ was increased.

\begin{figure}[h]
  \centering
  \includegraphics[width=\linewidth]{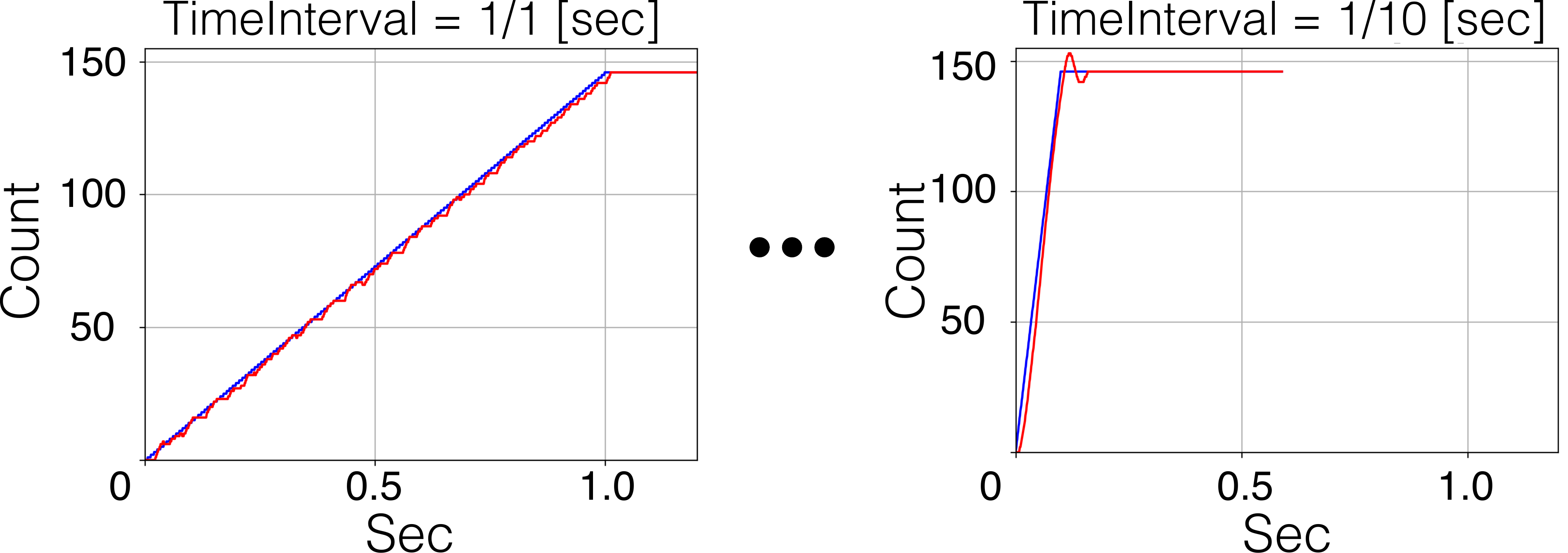}
  \caption{Experimental results of response speed of grass length control in \textit{ProgrammableGrass}.}
  \label{responsetime}
\end{figure}

Figure \ref{responsetime} shows the results of the response speed experiments. The blue and red lines are trajectories of the linear SPs and the PVs, respectively. The PVs were found to follow the linear SPs between 1 and  1/10  [sec]. Therefore, \textit{ProgrammableGrass} could control the grass length at a maximum of 10 [fps]. The response speed of \textit{ProgrammableGrass} was  10  times faster than the grass pixel response speed of 1 [sec] in the grass pixel method proposed by Tanaka et al. \cite{tanaka_dynamic_2023}. Furthermore, the response speed of  1/10  [sec] satisfies the speed of a limited animation, which is one of the animation methods to limit the number of frames per second to six or fewer \cite{whitaker_timing_2021}. Therefore, \textit{ProgrammableGrass} was found to be able to control the grass length enough to keep up with the animation speed. However, the grass pixel could not follow the linear SPs of  1/11  [sec] and shorter.

\vspace{-20pt}

\section{GRASS COLOR CALIBRATION SYSTEM}
\label{sectioncalib}

In this section, we focus on the ability to control the grass color based on the 8-bit levels in \textit{ProgrammableGrass}. When linear 8-bit levels are input to an LCD, the brightness changes nonlinearly due to a gamma characteristic. The 8-bit input level is then adjusted according to the gamma characteristic to control the brightness linearly according to the 8-bit levels. We apply this idea to \textit{ProgrammableGrass} to control the grass color linearly based on the 8-bit levels. 

The calibration of \textit{ProgrammableGrass} involves creating a correspondence table for each grass pixel to control the grass color at the 8-bit levels. This correspondence table maps the grass length to the 8-bit level and is derived from the relationship between the grass length and color in a grass pixel. We designed a grass color calibration system to obtain the correspondence table between the grass length and the 8-bit level. 

\subsection{Measurement of Grass Color}

\begin{figure*}[h]
  \centering
  \includegraphics[width=0.8\linewidth]{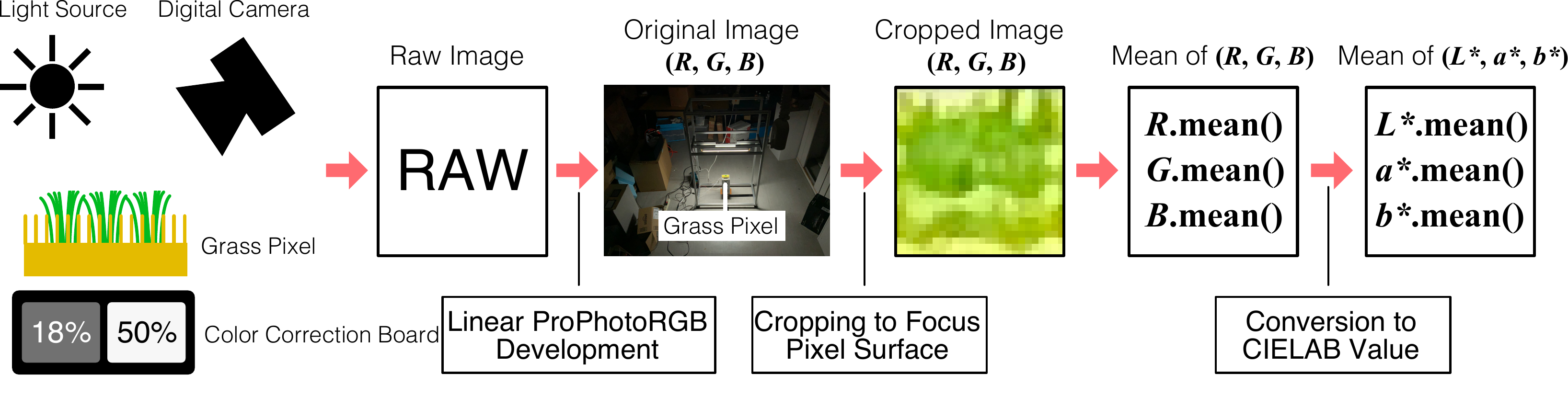}
  \caption{Measurement of grass color in CIELAB value through image processing.}
  \label{measurement}
\end{figure*}

In the grass color calibration system, the CIELAB color space is adopted to quantify the grass color because the CIELAB values of the grass colors can be compared using the color difference formula of the CIEDE2000. The CIEDE2000 is a color difference formula based on human color perception and is often used to measure multiple material colors, such as that of vegetables. In addition, each parameter of a CIELAB value is $L^*$ (Lightness), $a^*$ (Complementary colors between red and green), and $b^*$ (Complementary colors between yellow and blue). The CIELAB value of the grass color is called a grass color value in this study.

Figure \ref{measurement} shows the measurement of the grass color through image processing. A digital camera is used to capture the grass color. We prepare a color correction board including two types of gray cards, with 18\% and 50\% reflectance, respectively. The digital camera needs suitable exposure to capture the grass color stably using the color correction board. Before capturing the grass color, the camera's exposure is adjusted so that the intensity of the reflected light from the gray card with 18\% reflectance is in the middle of the light intensity range. 

The digital camera captures the grass pixel as a raw image to avoid a unique camera maker's image processing engine. The raw image is developed as an original image in the linear ProPhotoRGB color space. ProPhotoRGB is a device-independent RGB color space with a wide color gamut and is also used as a working color space in Adobe Lightroom \cite{adobe}. Since the relationship between light intensity and brightness is linear in the real world, a linear RGB color space of ProPhotoRGB is used to calculate the grass color. When the raw image is developed, the white balance is adjusted using the gray card with 50\% reflectance of the color correction board. After the raw image is developed, the original image is cropped to focus on the surface of the grass pixel.  Specifically, the three slits of the grass pixel fit into the cropped image. 

The grass color is quantified based on the cropped image. Since the grass color is perceived as the average color based on the spatial additive mixing, the average RGB value is calculated with the cropped image. The average RGB value is then converted to a CIELAB value via CIE 1931 color space. Since the white point of the linear ProPhotoRGB color space is 5000 [K], the white point of the CIELAB color space is also set to 5000 [K]. Therefore, the grass color value is calculated as a CIELAB value through image processing.

\subsection{Calculation of Correspondence Table}
\label{CalcCores}

\begin{figure}[h]
  \includegraphics[width=\linewidth]{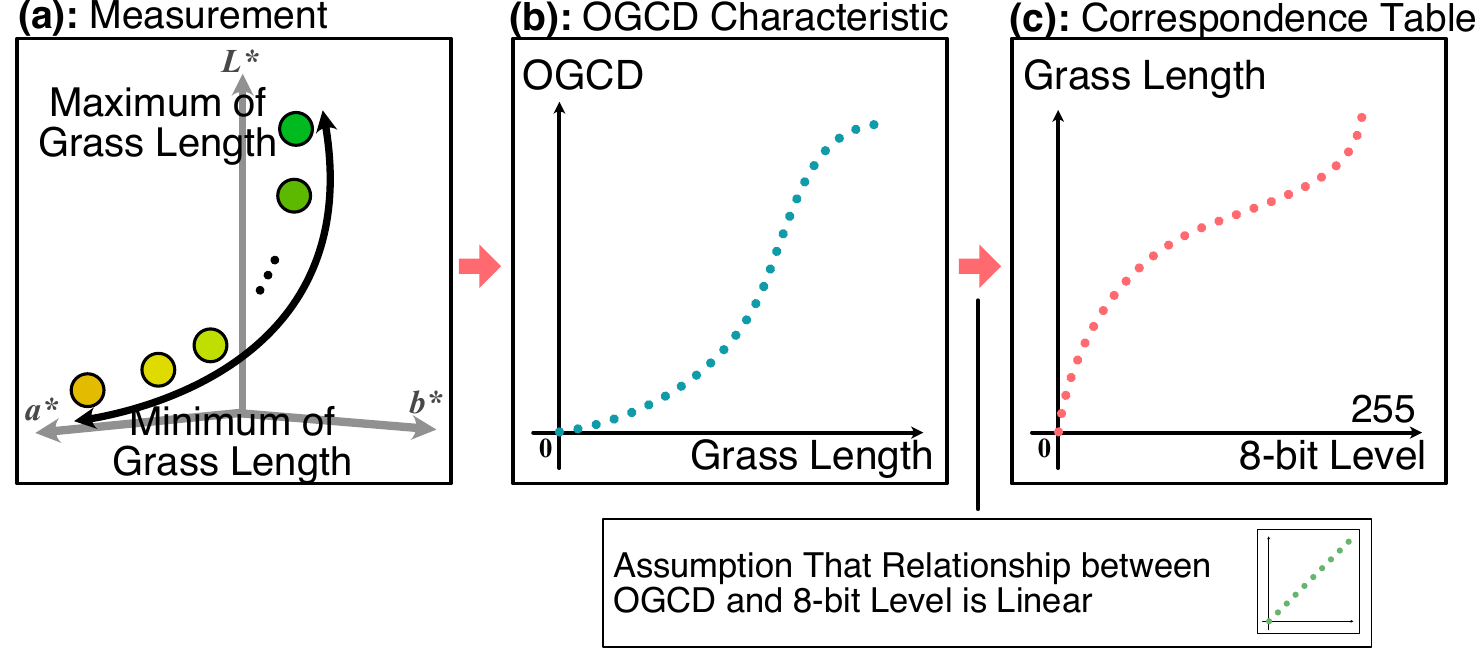}
  \caption{(a) Measurement of grass color values at constant grass length intervals, (b) OGCD characteristic calculated from measured grass color values using CIEDE2000 formula and (c) Calculation of Correspondence table between grass length and 8-bit level assuming linear relationship between OGCD and 8-bit level.}
  \label{process}
\end{figure}

In this subsection, we focus on how to obtain the correspondence table between the grass length and the 8-bit level using the measured grass color values. We describe the process of creating a correspondence table for a single grass pixel as shown in Figure \ref{process}.

First, the grass colors are measured at grass length intervals of about 1 [mm] between the minimum and maximum of the grass length, as shown in Figure \ref{process}(a). Second, the color difference of the grass color value at each grass length is calculated based on the grass color value of 0 [mm]. In this paper, OGCD (Origin Grass Color Difference) is defined as the color difference, quantified using the CIEDE2000 color difference formula in the CIELAB color space, between the grass color value at 0 [mm] and another measured grass color value. Therefore, the characteristic between the grass color and the grass length can be expressed by the OGCDs. Moreover, a nonlinear model is calculated by a sixth-order curve fitting using the calculated OGCDs between the minimum and maximum of the grass length, as shown in Figure \ref{process}(b). The nonlinear model is called an OGCD characteristic.

Finally, assuming a linear relationship between the OGCD and the 8-bit level where the maximum value of OGCD corresponds to 255 [level], a correspondence table between the grass length and the 8-bit level is obtained based on the OGCD characteristic model as shown in Figure \ref{process}(c). Therefore, the grass pixel can be calibrated when the obtained correspondence table is implemented in its software as described in Subsection \ref{sectionsoft}. Thus, \textit{ProgrammableGrass} can control the grass color linearly at the 8-bit levels.

\subsection{Calibration for Multiple Grass Pixels}
\label{CalibEach}

\begin{figure}[h]
  \includegraphics[width=\linewidth]{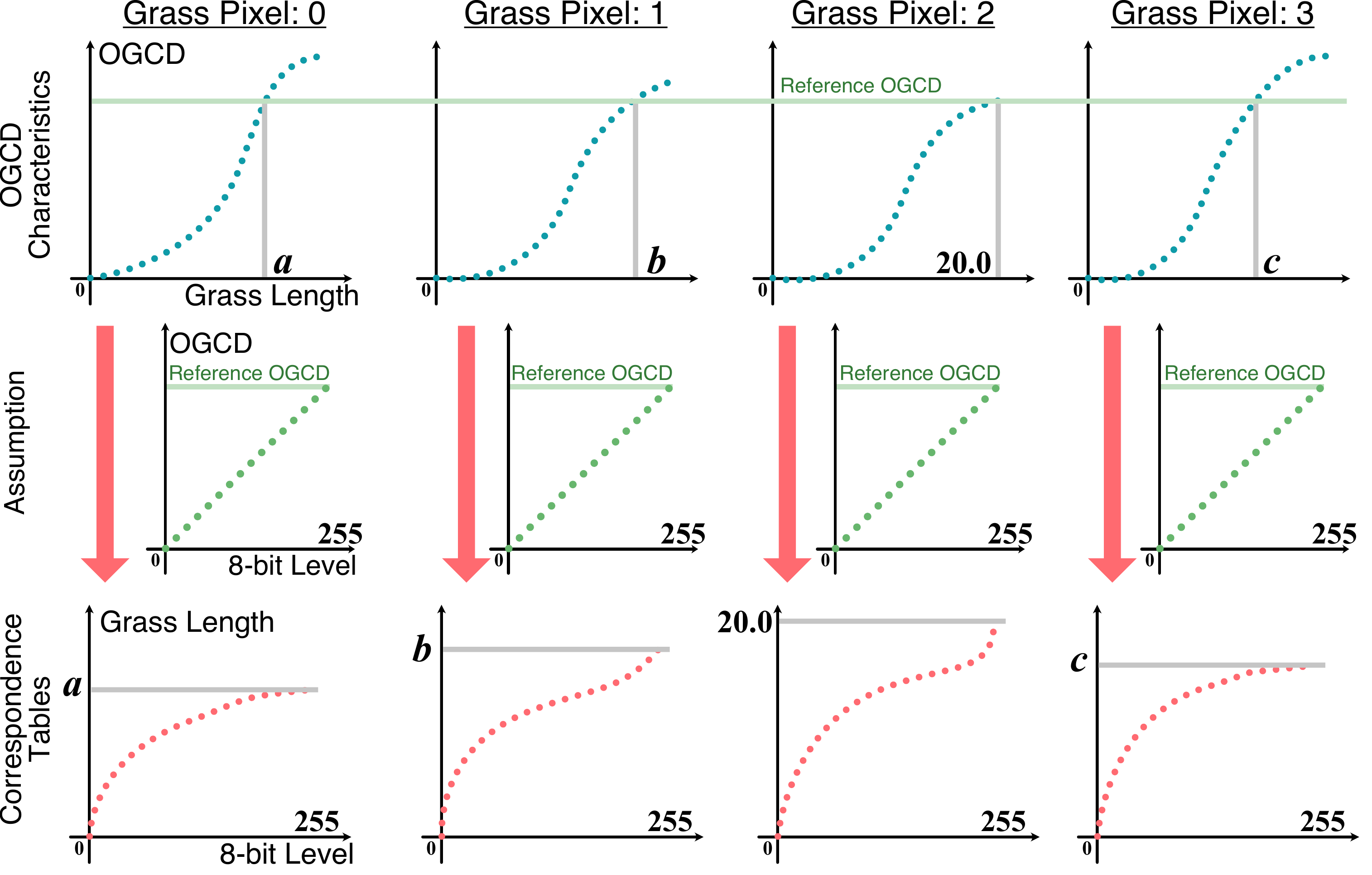}
  \caption{Overview of calibration based on multiple OGCD characteristics of grass pixels}
  \label{MultiCalibMethod}
\end{figure}

The grass pixels have different OGCD characteristics from pixel to pixel due to slight differences in green grass movement and the shape of the grass multi-slit component. Therefore, it is necessary to obtain correspondence tables that consider the OGCD characteristics of multiple grass pixels. This approach enables the control of the colors of multiple grass pixels at 8-bit levels, with minimal variation from pixel to pixel. Using the calibration of four grass pixels as an example, Figure  \ref{MultiCalibMethod} shows the output of the correspondence table for each grass pixel.

As outlined in Subsection \ref{CalcCores}, the OGCD characteristic for each grass pixel is obtained. Then, among the multiple OGCD characteristics,  the grass pixel with the smallest OGCD range is determined. The OGCD of this grass pixel at 20 [mm] grass length is called the reference OGCD.  Then, the correspondence table for each grass pixel is calculated so that the relationship between the OGCD and the 8-bit level is linear and the value of OGCD is the reference OGCD when the 8-bit level is 255.

For example, as illustrated in Figure \ref{MultiCalibMethod}, the reference OGCD is the OGCD at a grass length of 20 [mm] of the 2nd grass pixel, which has the smallest range of OGCD values. In the 0th grass pixel, the grass length $a$, which corresponds to this reference OGCD, is determined based on the OGCD characteristic. Then, the correspondence table is calculated where the grass length $a$ corresponds to an 8-bit level value of 255. Similarly, in the 1st and 3rd grass pixels, the correspondence tables are calculated using the grass lengths $b$ and $c$, respectively. These lengths correspond to the reference OGCD in each respective pixel, based on their unique OGCD characteristics.

Thus, by implementing these individual correspondence tables into each grass pixel, the calibration system accounts for the unique OGCD characteristics, providing a consistent method for controlling the color of multiple grass pixels at 8-bit levels.

\section{CALIBRATION TOOL USING TABLET COMPUTER}
\label{calibTool}

\begin{figure}[h]
  \includegraphics[width=\linewidth]{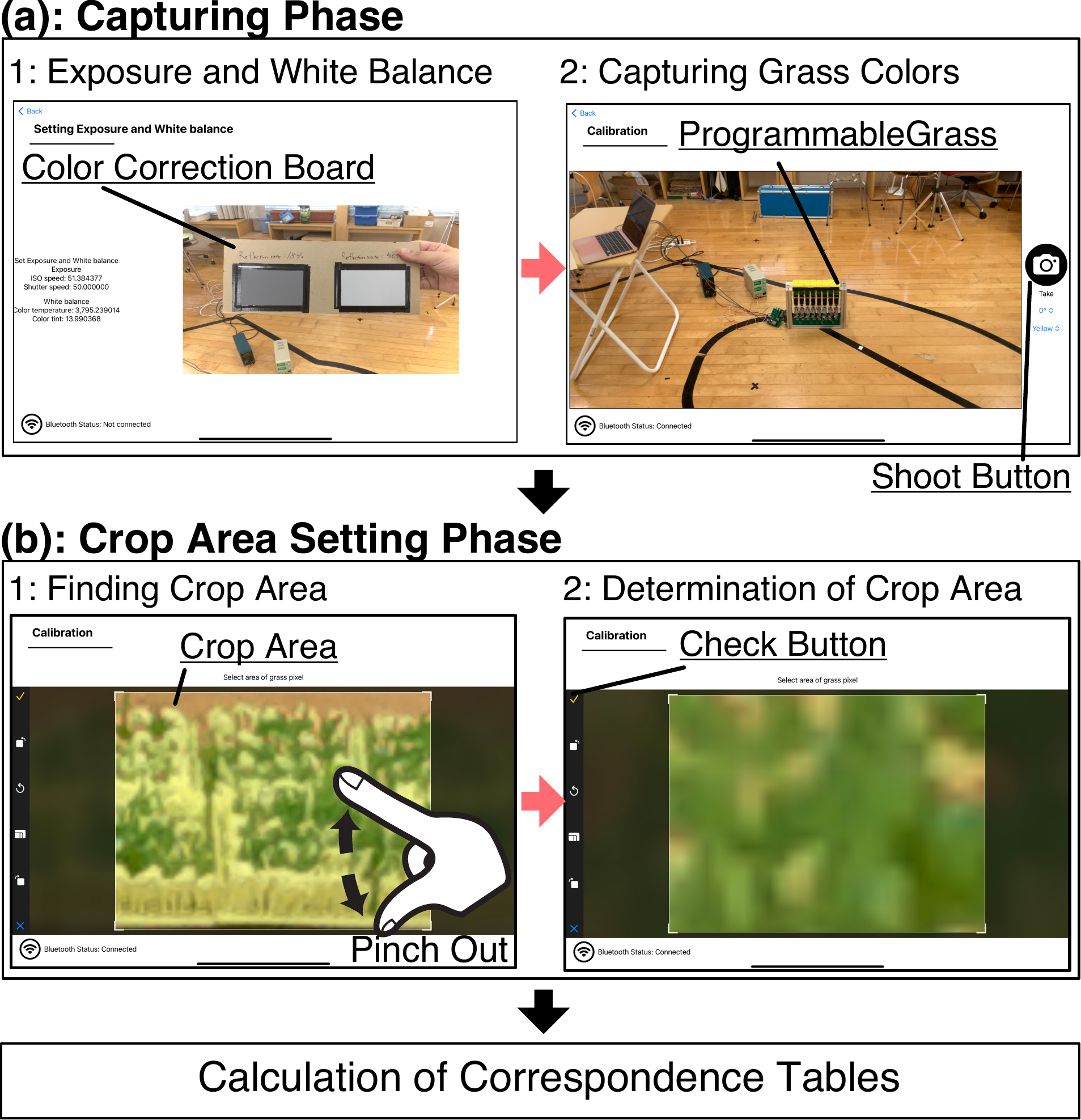}
  \caption{8-bit grass color calibration tool including (a) capturing phase, and (b) crop area setting phase}
  \label{iPad}
\end{figure}

We developed a calibration tool using a tablet computer to easily conduct the grass color calibration system. This tool was designed to manage three key aspects: capturing raw images of the colors of the grass pixels at grass-length regular intervals, determining the crop areas for each grass pixel within these images, and using these elements to create the correspondence tables for each pixel. In this paper, the calibration tool was developed based on our artificial grass display in Section \ref{HardModule}, and was used to calibrate the grass pixels in evaluations of the calibration system in Section \ref{SingleEval} and \ref{MultiEval}. We chose an iPad Pro (11-inch, 1st generation) as a digital camera (resolution: 12MP, aperture: f/1.8, lens: a wide lens) and a computer. Figure \ref{iPad} shows how to use the calibration.

\subsection{Capturing Phase}

In this phase, raw images are taken while moving the grass length from 0 to 20 [mm]. Before capturing, the exposure and white balance of the tablet's camera are adjusted using the color correction board as shown in Figure \ref{iPad}(a1). To move the grass length at regular intervals, the tablet and \textit{ProgrammableGrass} are connected through Bluetooth. The tablet is fixed at a certain position based on where a user looks at \textit{ProgrammableGrass}. As shown in Figure  \ref{iPad}(a2), the preview and the shoot button of the tablet's camera are displayed.   When the user pushes the shoot button, a raw image is captured each time the grass length is incrementally adjusted by approximately 1 mm, spanning the range from 0 to the maximum of 20 [mm]. At this time, the tablet computer sends the grass lengths as the SPs of the PD control instead of the 8-bit levels. 

\subsection{Crop Area Setting Phase} 

After the raw images are captured, a crop area is set for each grass pixel to measure the grass colors. The user slides or pinches the tablet screen to determine the crop area as shown in Figure \ref{iPad}(b1). Then, when the user pushes the check button, the crop area is decided as shown in Figure \ref{iPad}(b2). Since the tablet position is fixed, the crop area is applied to all raw images. 

Once the crop areas are determined, the CIELAB values of the grass colors for each grass length are obtained, and the correspondence table for each grass pixel can be calculated according to Section \ref{sectioncalib}. Thus, \textit{ProgrammableGrass} can be calibrated with the resulting correspondence tables.

\section{EVALUATION USING SINGLE GRASS PIXEL CALIBRATION}
\label{SingleEval}

We conducted indoor experiments to test whether \textit{ProgrammableGrass} can control the grass color linearly based on 8-bit levels using the grass color calibration tool. In this section, we measured and analyzed the color of a single calibrated grass pixel to confirm whether the grass color can be linearly and repeatedly controlled in the 8-bit levels. In the next section, we evaluated the effect of calibrating multiple grass pixels to reduce pixel-to-pixel color variation.

\subsection{Experimental environment}

\begin{figure}[h]
  \centering
  \includegraphics[width=0.7\linewidth]{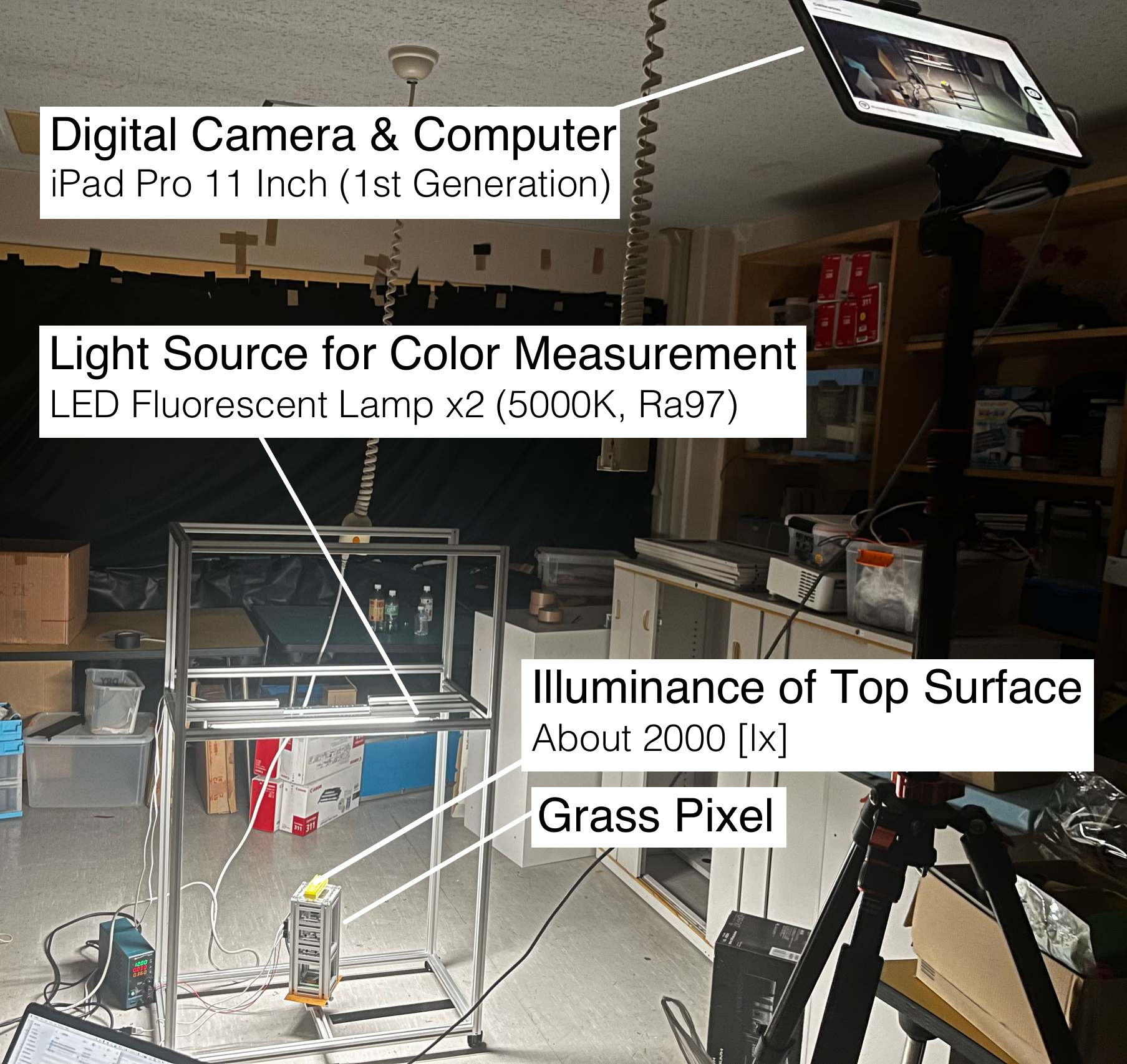}
    \caption{Experimental color measurement environment}
    \label{environment}
\end{figure}

Figure \ref{environment} shows the experimental environment. In the experiments, the grass pixel described in Subsection \ref{response} was used as \textit{ProgrammableGrass}. To measure the grass color stably, we built a color measurement environment focusing on International Standards Organization (ISO) 3664:2009 \cite{international_standard_organization_iso_iso_2009}, which is a standard of color measurements in the fields of graphic arts and photography where color management is a strict discipline. As shown in Figure \ref{environment}, the measurement environment was a dark room to avoid external lights. Two LED fluorescent lamps for color measurements were used as light sources (ECORICA LED, ECL-LD2EGN-L3A, color temperature: 5000 [K], average of color rendering index (Ra): 97). The light sources were placed at the height of 0.4 [m] from the top surface of the grass pixel so that the illuminance of the top surface was about 2000 [lx]. The color measurement environment with the light sources generally satisfies ISO 3664:2009. Silk gray cards with 18\% and 50\% reflectance were used as gray cards of the color correction board. An iPad Pro was used to measure the grass colors of the grass pixel and to use the calibration tool of Section \ref{calibTool}.

\begin{figure}[h]
  \centering
  \includegraphics[width=0.7\linewidth]{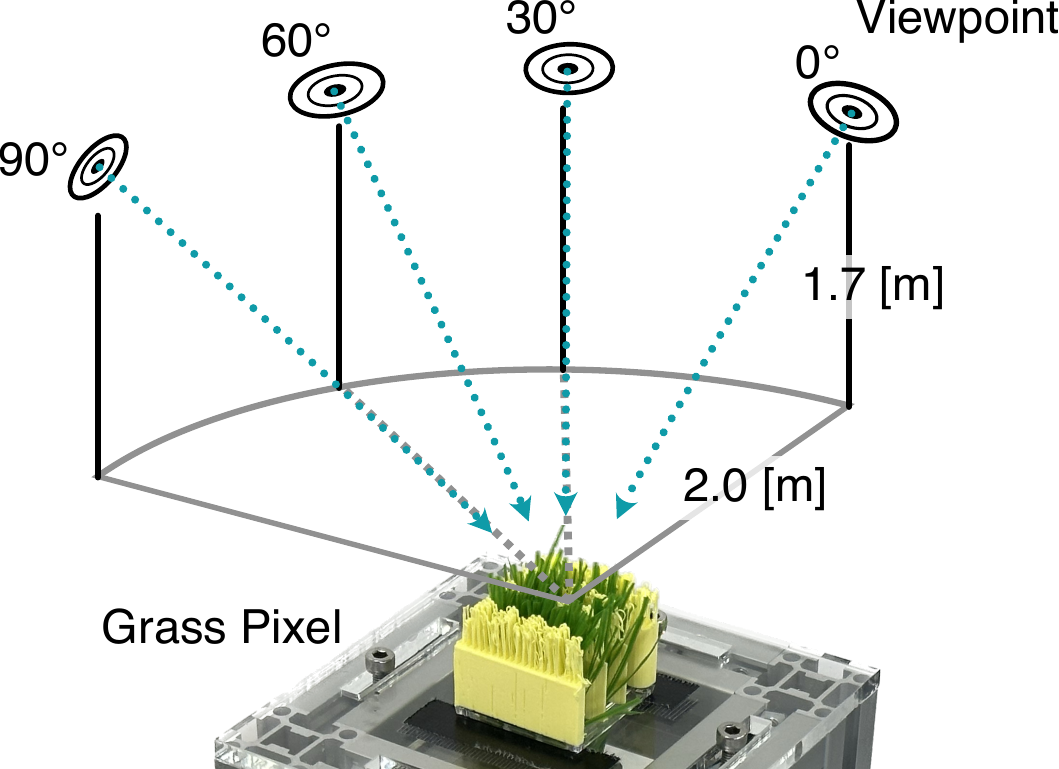}
    \caption{Four viewpoints based on horizontal angles with grass pixel in evaluation experiments}
    \label{viewpos}
\end{figure}

In the experiments, since the slits of the grass multi-slit component caused the grass pixel appearance differently depending on a viewer's position, the grass color was evaluated from several positions. Figure \ref{viewpos} shows the overview of the viewpoint positions. Since we assumed that \textit{ProgrammableGrass} was put down, the distance between the grass pixel and the viewpoint was 2.0 [m], and the height of the viewpoint was 1.7 [m] from the top surface of the grass pixel for the average adult height. The horizontal angles between the grass pixel and the viewpoint were $0^\circ, 30^\circ, 60^\circ$, and $90^\circ$ around the center of the grass pixel. When the horizontal angles were $0^\circ$ and $90^\circ$, it was perpendicular and parallel between the slits and the viewpoint, respectively. The iPad Pro captured the grass color at these viewpoints.

\subsection{Experimental Procedure}

Herein, the grass colors of the calibrated grass pixel were measured when the 8-bit levels were input at regular intervals. The grass pixel was calibrated at each viewpoint using the calibration tool in the experiments. In addition, the blue area of Figure \ref{8bitResult} shows the OGCD characteristic used to calibrate the grass pixel at each viewpoint. These OGCD characteristics of the blue curves illustrate the nonlinear characteristics between the grass color and the grass length for the grass pixel.

\subsubsection{Measurement Process for Calibrated Grass Pixel}
\label{sectionmeasurement}

The grass pixel was calibrated and measured from the $0^\circ, 30^\circ, 60^\circ$, and $90^\circ$ angles to evaluate the change in the grass color based on the 8-bit levels. The measurement process is described below. The red area of Figure \ref{analysis} shows the overview of the measurement method. 

\begin{enumerate}[i)]
  \item The grass pixel to be measured is selected from among those calibrated at the $0^\circ, 30^\circ, 60^\circ$, and $90^\circ$ viewpoints.
  \item The 8-bit levels were input into the selected calibrated grass pixel at interval 8 [level] at each viewpoint. In addition, the grass color values were measured each time. Specifically, the 8-bit input levels were 0, 7, 15, 24, ..., and 255 [level]. The 8-bit levels were increased by 8 intervals starting from 7 [level]. This process was considered to be one trial. 
  \item This trial was repeated 10 times, and the 8-bit input levels were raised and lowered five times each. We named the dataset of the grass color values for 10 trials as a measurement dataset. The measurement dataset was obtained from the $0^\circ, 30^\circ, 60^\circ$, and $90^\circ$ viewpoints.
\end{enumerate}

In the experiments, we used the above process to measure the grass colors of the grass pixel calibrated at $0^\circ, 30^\circ, 60^\circ$, and $90^\circ$. Therefore, 16 measurement datasets were obtained through the measurement process.

\subsubsection{Analysis Process of Measurement Dataset}

\vspace{-15pt}

\begin{figure}[h]
  \centering
  \includegraphics[width=\linewidth]{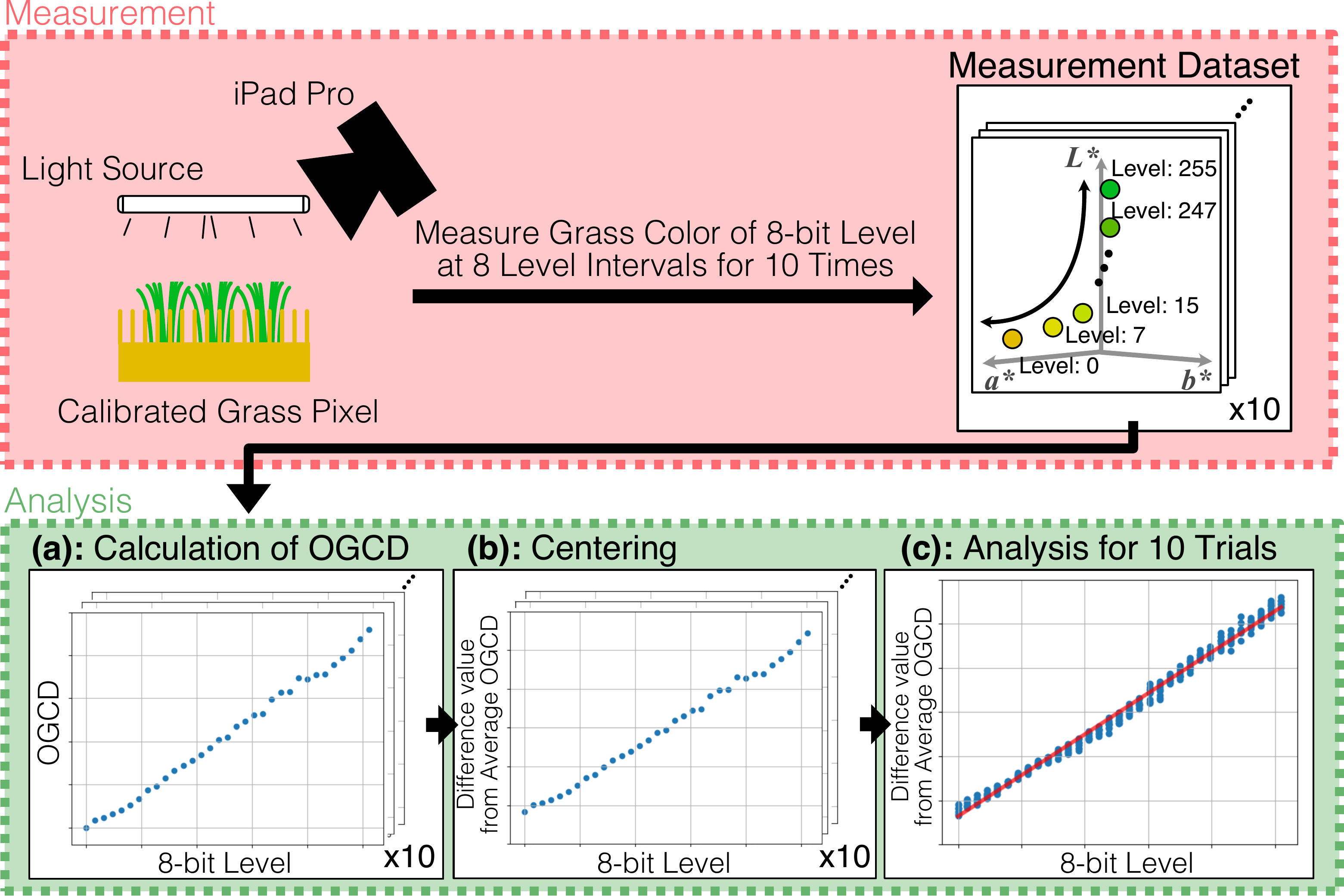}
  \caption{Measurement and analysis of experiments of grass color calibration system: (a) Calculation of OGCD based on 8-bit levels using measurement dataset for each trial, (b) Difference value of OGCD from average OGCD for each 8-bit level calculated from OGCD based on 8-bit levels for each trial and (c) Linear regression with difference values of OGCD from average OGCD for ten trials together.}
  \label{analysis}
\end{figure}

\begin{figure*}[h]
  \includegraphics[width=\linewidth]{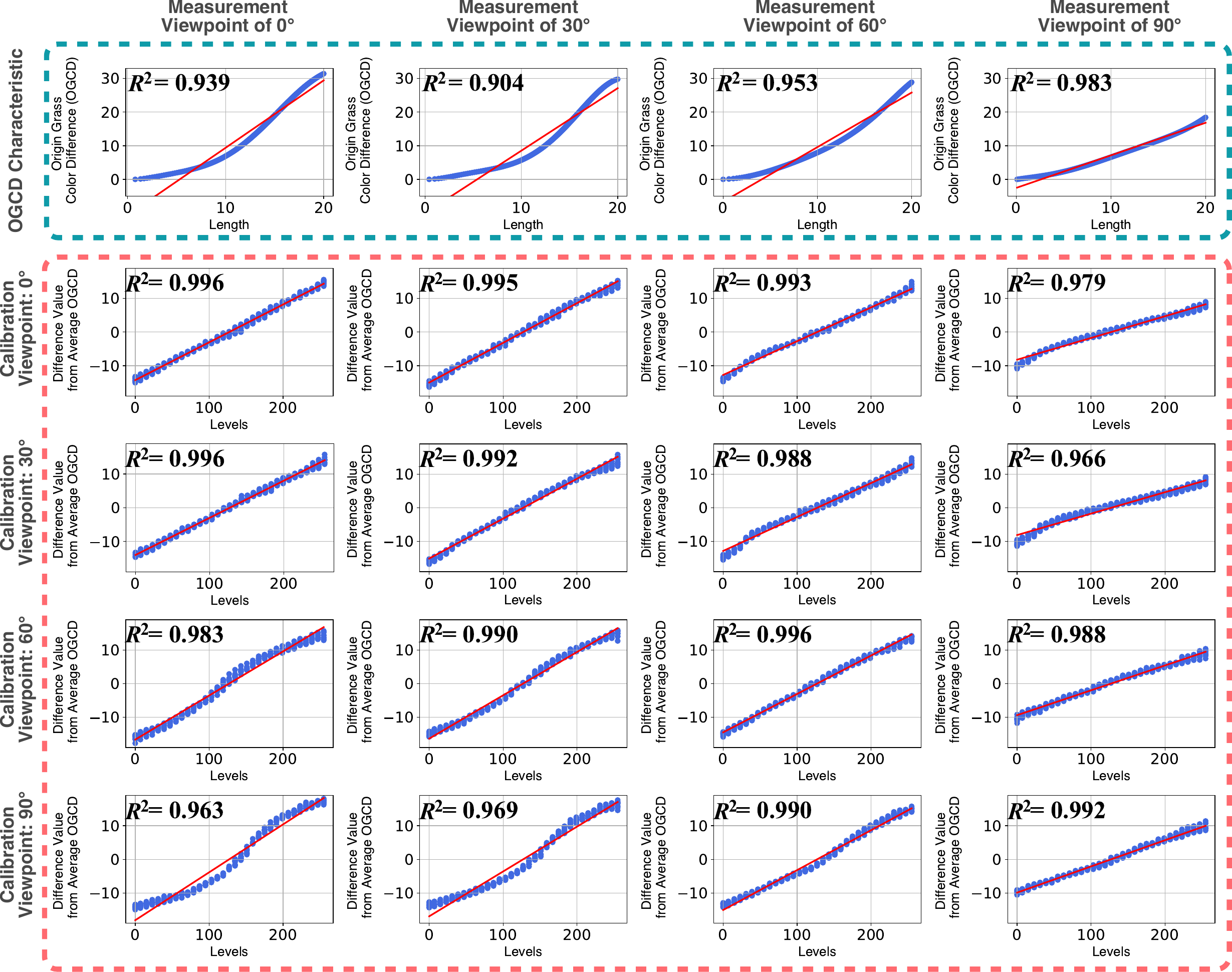}
  \caption{Experimental results of grass color calibration system with OGCD characteristics for each measurement viewpoint}
  \label{8bitResult}
\end{figure*}

Herein, we describe the analysis process using the measurement datasets to reveal whether the grass colors of the calibrated grass pixel can be linearly controlled based on the 8-bit levels. The analysis process is described below. The green area of Figure \ref{analysis} shows the analysis process overview.

\begin{enumerate}[i)]
  \item The OGCD for each 8-bit input level was calculated using the measured grass color values for each trial of the measurement dataset as shown in Figure \ref{analysis}(a). Since the 0 [level] corresponded to a grass length of 0 [mm], the OGCD for each trial was based on the measured grass color value of 0 [level].
  \item The difference value of the OGCD from the average OGCD for each trial was obtained for each 8-bit input level as shown in Figure \ref{analysis}(b).
  \item The single regression analysis was performed on the difference values of the OGCD from the average OGCD for ten trials together as shown in Figure \ref{analysis}(c). Then, the coefficient of determination $R^2$ and the graph, which includes the difference values of the OGCD and the predictive linear model, were obtained. The $R^2$ takes a value between 0 and 1. The closer $R^2$ is to 1, the higher the linearity of the analyzed dataset.
\end{enumerate}

Using this analysis process, we evaluated the change in the grass color of the grass pixel calibrated at $0^\circ, 30^\circ, 60^\circ$, and $90^\circ$ at each viewpoint position.

\subsection{Results and discussion}

The red area of Figure \ref{8bitResult} shows the experimental results of the single calibrated grass pixel. In the experimental results, the viewpoints from which the grass pixel was calibrated and measured are referred to as calibration viewpoints and measurement viewpoints, respectively. The rows and columns are classified by the calibration viewpoints and the measurement viewpoints, respectively. For the experimental results, the blue dots show the difference values of the OGCD from the average OGCD for each 8-bit level of the ten trials together. The red line is the linear model calculated from the single regression analysis based on the difference values of the blue dots. In addition, the $R^2$ is shown in each experimental result. The blue area in Figure \ref{8bitResult} also shows the OGCD characteristics for comparison with the experimental results. The OGCD characteristics represent the nonlinear characteristics between the grass length and color of the grass pixel. In addition, the OGCD characteristics were also evaluated with a single regression analysis to show the linear model as the red line and the $R^2$.

\begin{figure}[h]
  \includegraphics[width=\linewidth]{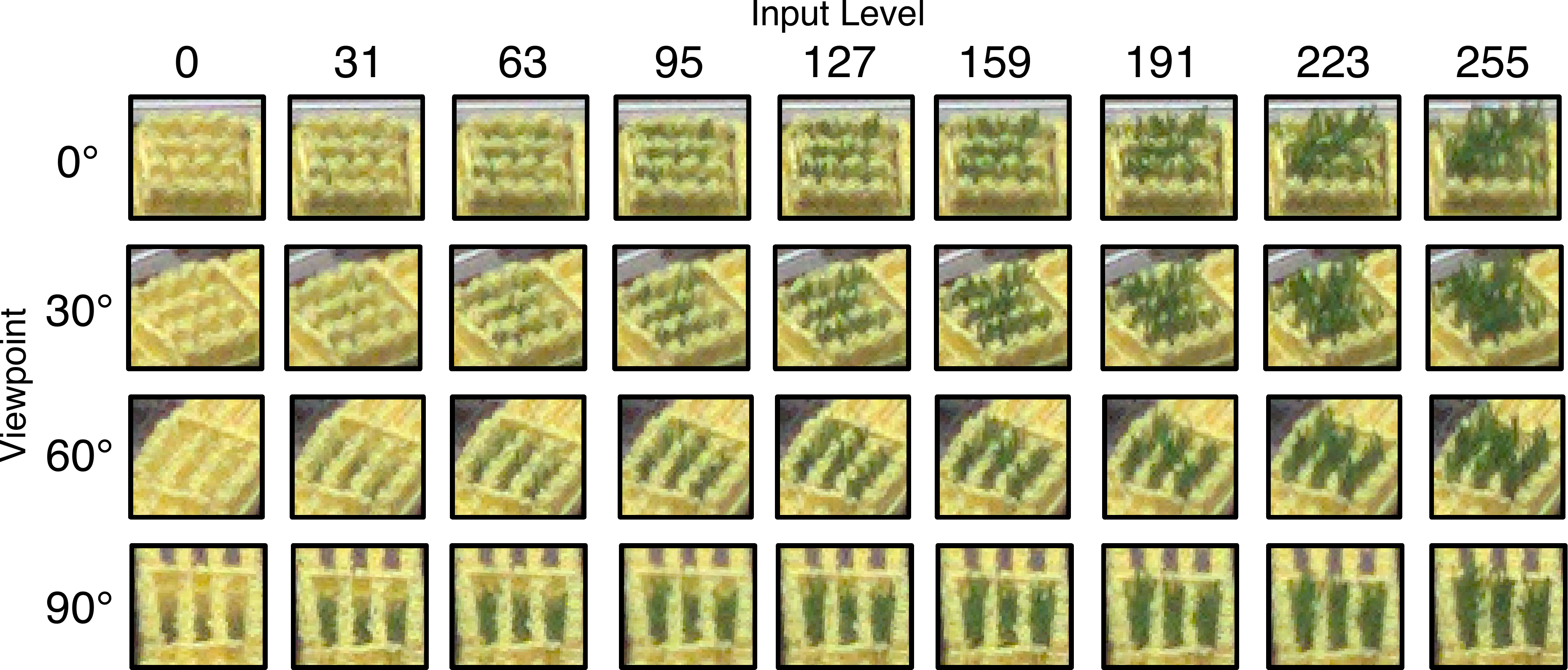}
  \caption{Experimental visual results of the grass pixel calibrated $0^\circ$ from four viewpoints}
  \label{VisualResult}
  \vspace{-15pt}
\end{figure}

As a result, we revealed that the grass color of the grass pixel could be controlled linearly based on the 8-bit levels using the grass color calibration system. Additionally, in all experimental results, the $R^2$ values were greater than 0.9, indicating that the grass pixel could consistently control the grass color across all viewpoints. In the results at the $0^\circ$ calibration viewpoint, the blue dots were plotted linearly at each measurement viewpoint. In particular, the experimental graph of the $0^\circ$ measurement viewpoint shows that the strong nonlinear OGCD characteristic at the $0^\circ$ measurement viewpoint was improved. In addition, the $R^2$ of the OGCD characteristic improved from  0.939  to  0.996. In the result at even the $30^\circ$  and $60^\circ$  measurement viewpoint, the $R^2$ values of the OGCD characteristic improved from  0.904  to  0.995 and from 0.953 to 0.993, respectively. At the $90^\circ$ measurement viewpoints, the $R^2$ of the OGCD characteristics did not improve. However, the plot results show that the grass color changed linearly. Figure \ref{VisualResult} shows a set of images of the grass pixel, calibrated at the $0^\circ$ viewpoint and captured from four different viewpoints. 

 Similar to the results of the $0^\circ$ calibration viewpoints, the calibration at the $30^\circ$ improved the $R^2$ of the OGCD characteristic at $0^\circ$, $30^\circ$, and $60^\circ$ measurement viewpoints except for $90^\circ$ measurement viewpoint. Moreover, the $R^2$ at all measurement viewpoints improved at the $60^\circ$ and $90^\circ$ calibration viewpoints. 

Furthermore, we also revealed that the effectiveness of the grass color calibration system depended on the calibration viewpoints. At the $0^\circ$  and $30^\circ$  measurement viewpoint, the OGCD characteristic was nonlinear with the grass length. Then, the OGCD characteristic began to change linearly from the $60^\circ$ measurement viewpoint.  This was caused by the slit appearance of the grass pixel.  At the $60^\circ$ and $90^\circ$ measurement viewpoints,  the green grass was seen through the slits of the grass pixel earlier than at the $0^\circ$ and $30^\circ$ measurement viewpoints.  Therefore, since the grass pixel was calibrated with the OGCD characteristic, the effectiveness of the grass color calibration system depended on the calibration viewpoints. 

For example, when the grass pixel calibrated at the $0^\circ$ viewpoint was measured from the $60^\circ$ viewpoint, the plotted graph was more convex above the OGCD characteristic at the $60^\circ$ measurement viewpoint. This was due to the strong effect of linearizing the nonlinear OGCD characteristic at the $0^\circ$ measurement viewpoint. In contrast, when the grass pixel calibrated at the $60^\circ$ viewpoint was measured at the $0^\circ$ viewpoint, the grass color changed linearly; however, the grass pixel was not improved as much as the grass pixel calibrated at the $0^\circ$ viewpoint. This was because the effect of the grass color calibration system of the $60^\circ$ viewpoint was weak in controlling the grass color observed from the $0^\circ$ viewpoint.

The experimental results show that the range of the OGCD of the $90^\circ$ measurement viewpoint was smaller than that of the other measurement viewpoints. This was because the green grass was seen even though the grass length was 0 [mm] at the $90^\circ$ measurement viewpoint. This phenomenon did not occur from other measurement viewpoints. Therefore, the amount of the change in the OGCD of the $90^\circ$ measurement viewpoint was smaller than that of other measurement viewpoints.

We obtained the average value of the $R^2$ from the $0^\circ$ to $90^\circ$ measurement viewpoints, in order of the $0^\circ, 30^\circ, 60^\circ$, and $90^\circ$ calibration viewpoints:  0.9908, 0.9855, 0.9893, and 0.9785,  respectively.  The average $R^2$ value was highest at the $0^\circ$ calibration viewpoint. Therefore, we concluded that the $0^\circ$ viewpoint is the most suitable calibration viewpoint when \textit{ProgrammableGrass} is observed from angles ranging from $0^\circ$ to $90^\circ$. 

\section{CALIBRATED GRASS PIXEL EVALUATION IN SEVERAL ENVIRONMENTS}

\subsection{EXPERIMENTAL PROCEDURE}

In the previous section, we conducted experiments on the calibration effects of the grass pixel under the standard lighting conditions of ISO with high luminance, high color rendering, and a color temperature of 5000K. In this section, we delve into the calibration effects within actual indoor environments by undertaking experiments in several indoor settings.

The study was carried out in three distinct locales within the University of Tsukuba campus: a classroom, a meeting room, and a dimly lit space. These experiments employed three viewpoints: $0^\circ, 30^\circ$ and $60^\circ$ as discussed in Section \ref{SingleEval}. Here, we installed the grass pixel and performed calibration at $0^\circ$ in each indoor environment. Subsequently, we input levels ranging from 0 to 255 at fixed intervals into the grass pixel and measured the OGCD with the viewpoints $0^\circ, 30^\circ$ and $60^\circ$. This procedure was aimed at ascertaining whether the OGCD changes follow a linear pattern as the environment changes. Details of the illuminance and the color temperature at the location of the grass pixel in the experimental environments are summarized in Table \ref{ill}.

\begin{table}[h]
  \centering
  \caption{Details of illuminance and color temperature of each environment}
  \begin{tabular}{|c|c|c|}
  \hline
  \multicolumn{1}{|l|}{} & Illuminance & Color Temperature \\ \hline
  Classroom              & 404 lx      & 3564K             \\ \hline
  Meeting Room           & 525 lx      & 4380K             \\ \hline
  Dimly Lit Space        & 113 lx      & 4059K             \\ \hline
  ISO Environmet  (Reference)         & 2000 lx     & 5000K             \\ \hline
  \end{tabular}
  \label{ill}
\end{table}

\begin{figure*}[h]
  \includegraphics[width=0.9\linewidth]{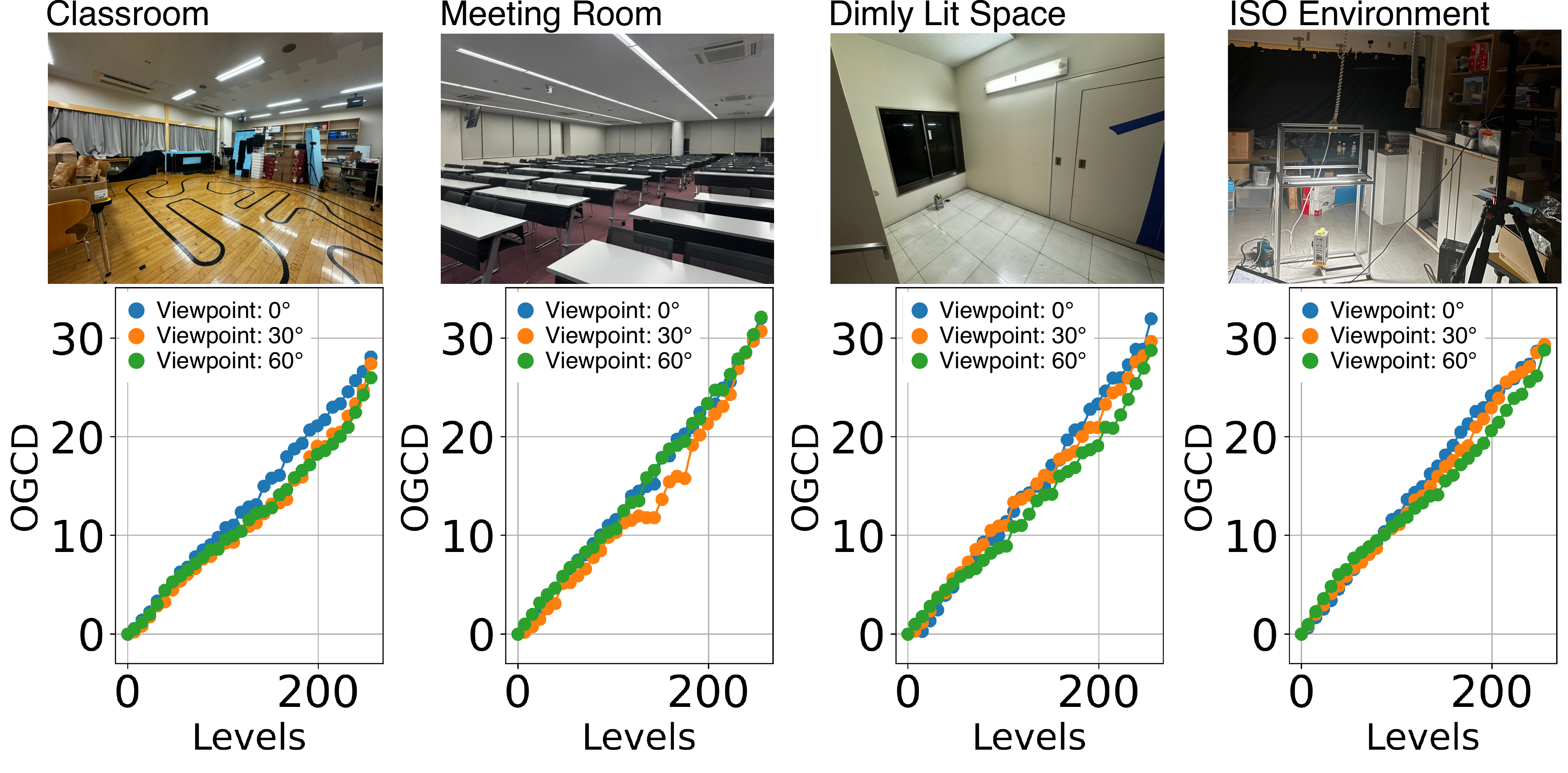}
  \centering
  \caption{Experimental results of the grass pixel calibrated at viewpoint $0^\circ$ in several environments}
  \label{multiEnv1}
\end{figure*}

\begin{figure*}[h]
  \centering
  \includegraphics[width=0.9\linewidth]{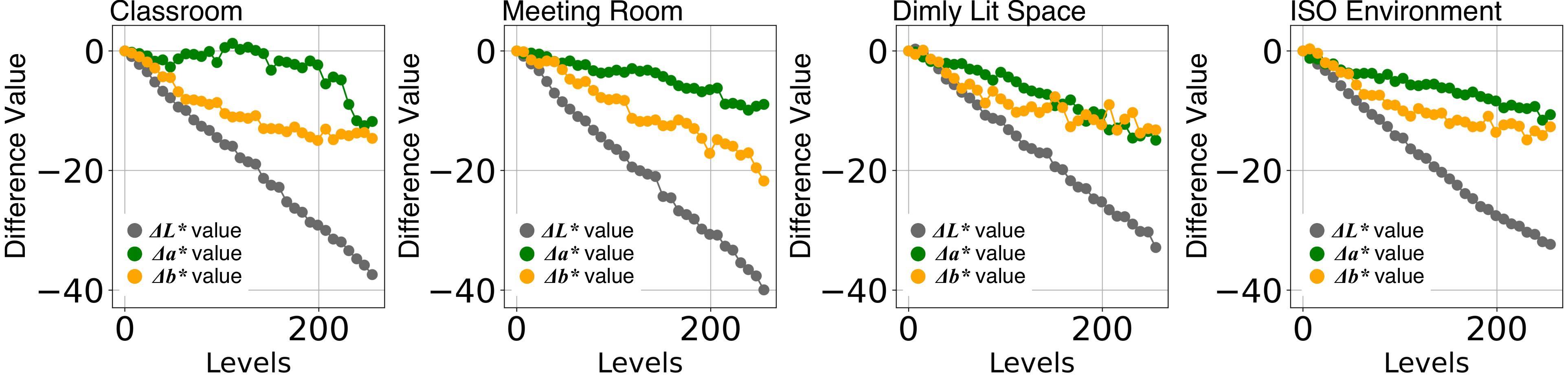}
  \caption{Changes of each $L^*, a^*$ and $b^*$ of the calibrated grass pixel in several environments}
  \label{multiEnv2}
\end{figure*}

\subsection{Results and discussion}

The experimental outcomes are depicted in Figure \ref{multiEnv1}. These illustrations not only incorporate the three indoor environments but also the results from the previous section's ISO standard environment experiments. As a result, we revealed that calibrating the grass pixel in three different indoor settings, similar to the ISO environment, allows for a linear control of the grass color in response to the input from the 8-bit levels. These color controls are consistent across the three viewpoints. Additionally, despite the changes in luminance and color temperature across the four environments presented in these graphs, the range of OGCD was found to be approximately 30.

The uniformity in the OGCD range across different environments can be attributed to the calibration process, where exposure and white balance were adjusted using 18\% and 50\% gray cards, respectively. This adjustment ensures that the changes in the grass color's $L^*, a^*$, and $b^*$ components in relation to the 8-bit levels are similar. To facilitate discussion on these results, Figure \ref{multiEnv2} showcases the changes in $L^*, a^*$, and $b^*$ observed from the viewpoint $0^\circ$ in each environment. The graph plots the differences from the 0 level values of $L^*, a^*$, and $b^*$, color-coded in gray, green, and yellow, respectively. The plots reveal that, across all environments, the range of differences in $a^*$ and $b^*$ is around 15, whereas for $L^*$, it's about 35. In these experiments, even in the dimly lit space with only 113 [lx], the change in $L^*$ was similar to that in the other three environments. However, we suppose that in environments with even lower illumination, despite adjusting the exposure of the digital camera, the color of the grass tends to be measured as darker, almost black. This, in turn, would lead to a smaller range of the OGCD.

\section{EVALUATION USING MULTIPLE GRASS PIXELS CALIBRATION}
\label{MultiEval}

Previously, we revealed that the single grass pixel can control its grass colors in the 8-bit levels using our grass color calibration system. In this section, we evaluated the ability of the calibration system to focus on multiple grass pixels.

\subsection{Experimental Procedure}

The experimental environment and procedure are similar to the experiments in Section \ref{SingleEval}. We measured the grass colors in the same color measurement environment as used in Section \ref{SingleEval}. A grass module was used as \textit{ProgrammableGrass} to be measured. The iPad Pro of the previous section was also used to measure the grass colors and to calibrate the grass pixels with the calibration tool. The iPad Pro was located a distance of 2.0 [m] from the grass module and at a height of 1.7 [m] above the surface of the grass module.

In these experiments, we evaluated the effect of calibrating the eight pixels of the grass module based on their respective OGCD characteristics. In Section \ref{SingleEval}, we revealed that it is suitable to calibrate \textit{ProgrammableGrass} from a $0^\circ$ viewpoint where the camera and the slits are perpendicular. Thus, the eight grass pixels were also calibrated from the $0^\circ$ viewpoint using the tablet calibration tool. The calibration generated multiple correspondence tables for the eight grass pixels. After calibration, as in the experiments in Section \ref{SingleEval}, 8 level intervals were input to the grass pixels and the OGCD of each grass pixel was measured each time from the $0^\circ$ viewpoint. We also measured the changes in OGCD for each grass pixel when applying a single correspondence table simultaneously to all eight pixels. In this procedure, every grass pixel employed the correspondence table derived from the grass pixel that demonstrated the smallest OGCD range during the calibration of multiple grass pixels. This ensured that the range of grass length in the single correspondence table was from 0 to 20 [mm].

\subsection{Results and discussion}

\begin{figure}[h]
  \includegraphics[width=\linewidth]{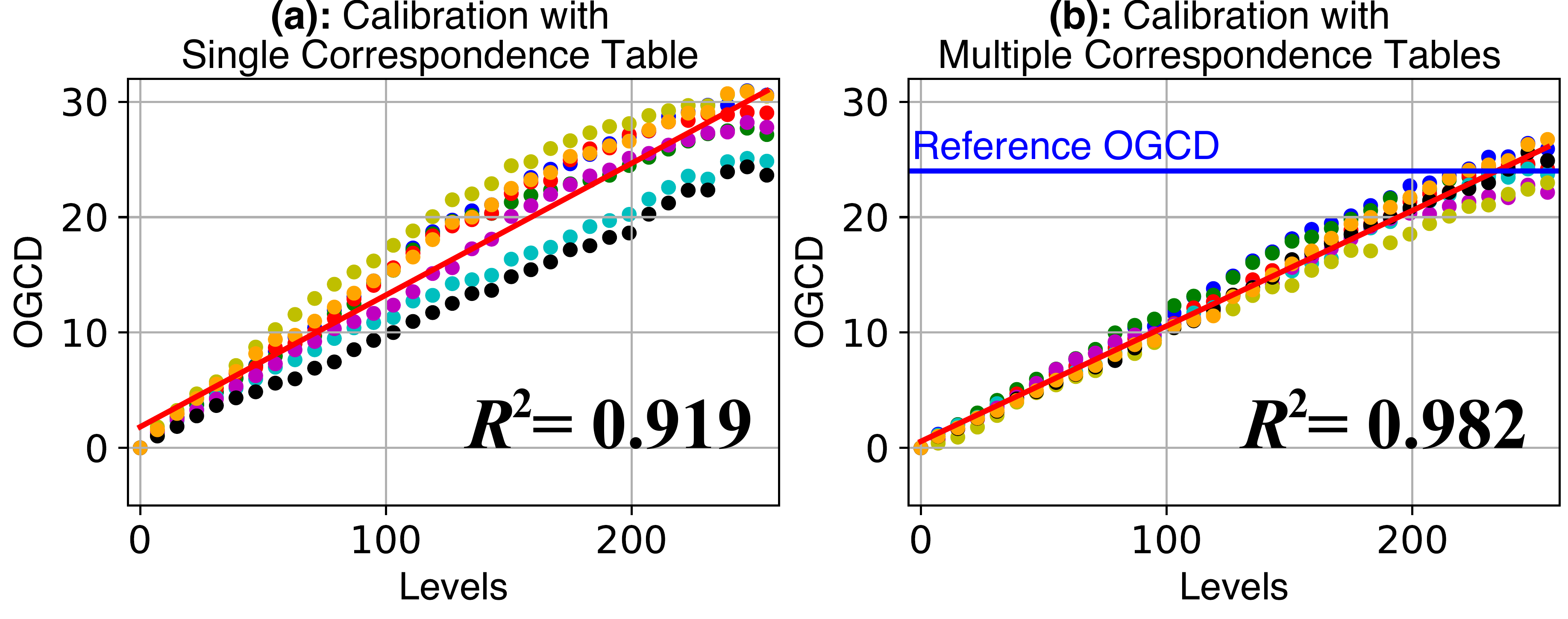}
  \caption{Experimental results of calibration with (a) a single correspondence table and (b) multiple correspondence tables}
  \label{MultipleCalib}
\end{figure}

\begin{figure*}[h]
  \centering
  \includegraphics[width=\linewidth]{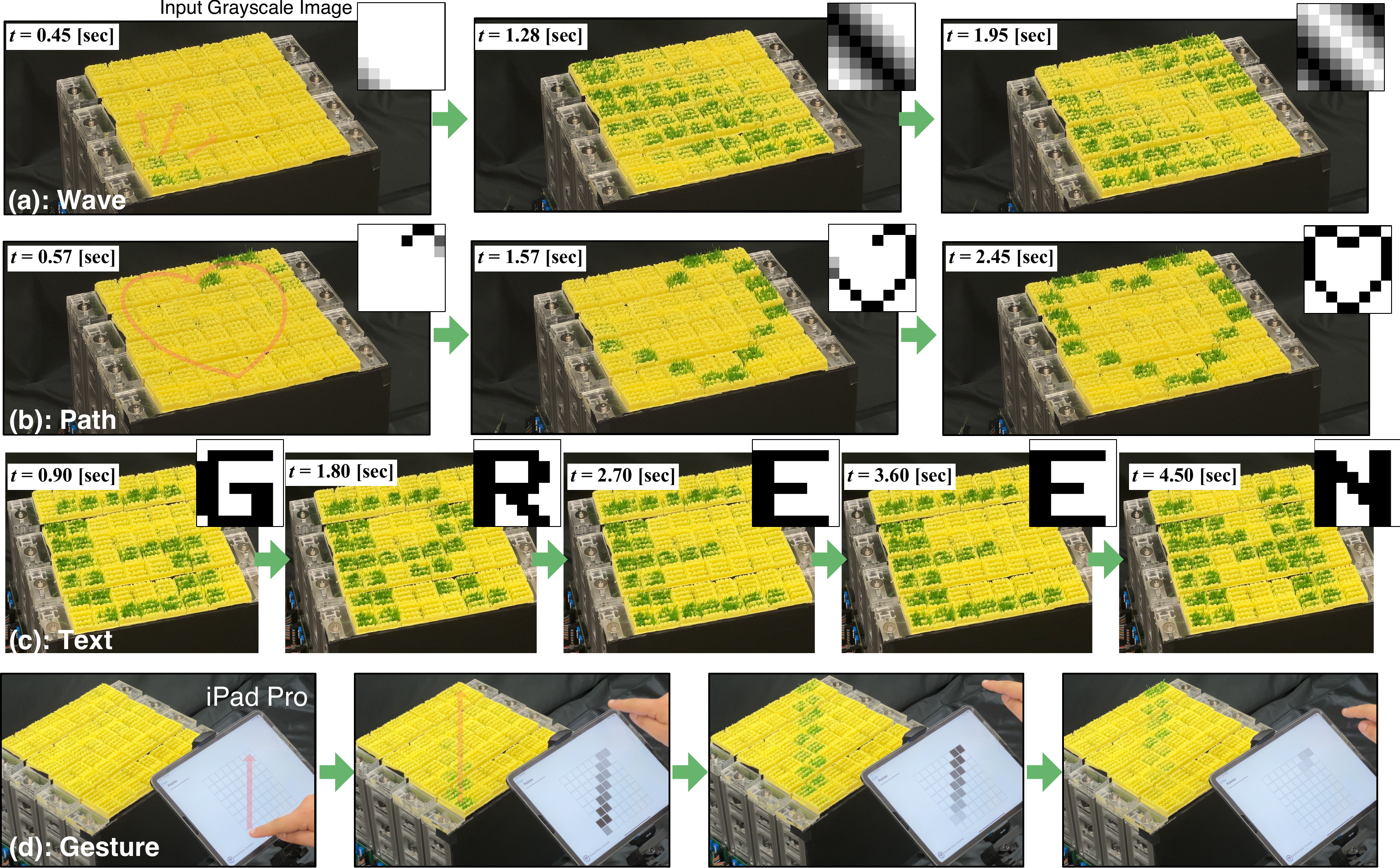}
  \caption{Animation examples: (a) Wave animation, (b) Path animation of heart mark, (c) Text animation of "GREEN", and (d) Gesture-based animation with a tablet }
  \label{animation}
\end{figure*}

Figure \ref{MultipleCalib}(a) and (b) show the results with the single correspondence table and multiple correspondence tables, respectively. The colors of the dots on the graphs are differentiated for each grass pixel. When the color of the dots is the same in graphs (a) and (b), it indicates that these dots correspond to the same grass pixel. In Figure \ref{MultipleCalib}(b), the blue line represents the reference OGCD, which was described in Subsection \ref{CalibEach}. Moreover, a single regression analysis was performed on Figure \ref{MultipleCalib}(a) and (b) to determine how similar the OGCD changes were for the eight grass pixels. Then, Figure \ref{MultipleCalib} shows the prediction model of the single regression analysis with a red line, and the coefficient of determination $R^2$ is also displayed. 

As a result, by creating correspondence tables for the eight grass pixels, it became easier to control the colors of the eight grass pixels. In the plot of (b), the OGCD variation among the grass pixels is smaller than in the plot of (a), and the $R^2$ value is improved from 0.919 to 0.982.

For the results in (a), the OGCD for black and cyan dots, which have small OGCD ranges, changes linearly, while other color dots have upward convex trajectories. This is because each grass pixel used the correspondence table for the grass pixel with small OGCD ranges. The largest OGCD difference between the grass pixels was 9.8 at level value 167. 

In contrast, in results (b), each grass pixel was calibrated based on its characteristics. Thus the grass colors of all pixels could be controlled linearly. In these experiments, the reference OGCD was set at 24.5. At level 255, the mean OGCD value was 24.3, with a standard deviation of 1.41. 

Thus, we revealed that the calibration based on multiple OGCD characteristics was effective in reducing the OGCD variation among the grass pixels.

\section{DEMONSTRATIONS}
\label{demo}

\subsection{Animation Examples}
\label{sectionanimation}

\begin{figure*}[h]
  \centering
  \includegraphics[width=0.8\linewidth]{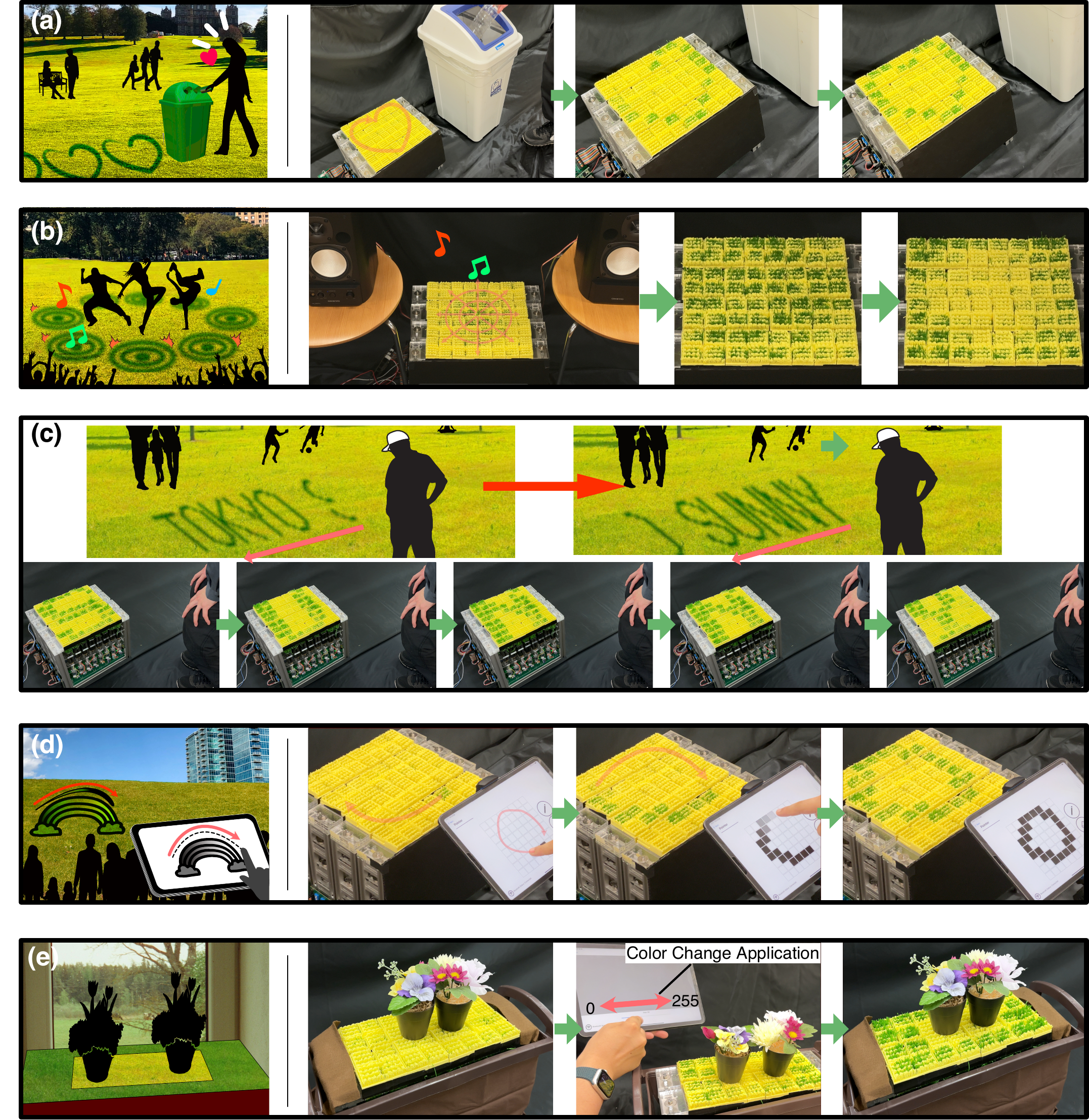}
  \caption{Application examples: (a) Trash thanker application, (b) Music visualizer, (c) Green signboard application, (d)  Green drawing application, and (e) Planter space for flowers}
  \label{application}
\end{figure*}

Here, we demonstrated several animation examples of \textit{ProgrammableGrass} as shown in Figure \ref{animation}. An $8\times8$ pixel artificial grass display was built using four grass modules in the animation examples.  In the examples of Figure \ref{animation}(a), (b), and (c), the artificial grass display played the animations at 10 [fps]  using several $8\times8$ MP4 files with 8-bit grayscale levels.  In addition, we calibrated the artificial grass display using the calibration tool in the demonstration environment 

\vspace{-5pt}

\subsubsection{Wave Animation}
As shown in Figure \ref{animation}(a), the artificial grass display presents a wave animation, demonstrating the smooth and dynamic grass color control based on the 8-bit levels. The wave animation showed the waves flowing from the lower left of the artificial grass display to the upper right. This demonstration shows that \textit{ProgrammableGrass} is capable of expressing dynamic motion with gradations through artificial grass materials.

\vspace{-5pt}

\subsubsection{Path Animation}
The artificial grass display demonstrated a path animation that draws a heart as shown in Figure \ref{animation}(b). Since the artificial grass display had $8\times8$ pixels and a quick response speed of 10 [fps], the graphical icon of the heart mark could be displayed with the fast and smooth path animation. 

\subsubsection{Text Animation}
As shown in Figure \ref{animation}(c), the artificial grass display, with a resolution capable of displaying alphabetic characters, demonstrated a text animation. The displayed text was "GREEN", and the text animation flowed from right to left smoothly. This demonstration shows the possibility of providing people with familiar information in their daily lives such as news and weather forecasts through artificial grass materials.

\vspace{-5pt}

\subsubsection{Gesture-based Animation}
Figure \ref{animation}(d) illustrates how the artificial grass display can present a gesture-based animation via a tablet computer through Bluetooth communication. The tablet application consisted of $8\times8$ pixel squares, and each square corresponded to a grass pixel of the artificial grass display. When the user touches a square, the 8-bit grayscale of the square changes linearly, and the grass color is controlled accordingly. As shown in Figure \ref{animation}(d), the fast 8-bit grass color control allows the artificial grass display to play the animation on the tablet when the user swiped from the bottom left to the top right. This animation presents the possibility of real-time interactive applications between the user and the artificial grass display.

\vspace{-5pt}

\subsection{Application Examples}
\label{SectionApp}

Based on the findings in the previous subsection, we made several applications of \textit{ProgrammableGrass} as shown in Figure \ref{application}.  To emphasize the practicality of these applications, we also included conceptual images in each application, illustrating the potential of these applications when \textit{ProgrammableGrass} could be integrated into society as if it were conventional grass.

\subsubsection{Trash Thanker Application}

\textit{ProgrammableGrass} can represent an emotion by displaying an animation of a simple icon as described in Subsection \ref{sectionanimation}. We made a trash thanker application through the path animation of the heart mark as shown in Figure \ref{application}(a). A recycling bin for plastic bottles is placed near \textit{ProgrammableGrass}. When a user discards a plastic bottle into the recycling bin, \textit{ProgrammableGrass} expresses gratitude by displaying a heart mark. Through this application, users can gain a sense of contributing to societal recycling efforts by discarding their plastic bottles in the correct location.

\subsubsection{Music Visualizer}

\textit{ProgrammableGrass} can express dynamic motions of gradations based on the 8-bit levels. As shown in Figure \ref{application}(b), we created a music visualizer through artificial grass materials. In this application, dance music at 100 beats per minute (BPM) is played, and \textit{ProgrammableGrass} displays a motion animation that pulses in sync with the beat. Through this application, art methods using grass as a medium are expanded with the addition of the animation techniques like as dance effects in grass fields. 

\subsubsection{Green Signboard Application}

Subsection \ref{sectionanimation} shows that \textit{ProgrammableGrass} can show the text animation smoothly. This feature allows \textit{ProgrammableGrass} to communicate text-based information to the users through artificial grass materials. Using the text animation, we created a green signboard application of \textit{ProgrammableGrass}. For example, \textit{ProgrammableGrass} can provide weather forecast information to the user, as shown in Figure \ref{application}(c).

\subsubsection{ Green Drawing Application }

Since \textit{ProgrammableGrass} can show the user's gesture animations via the tablet, we created a green drawing application of \textit{ProgrammableGrass}. In Figure \ref{application}(d), the user draws a circle on $8\times8$ squares on the tablet computer. Then, the artificial grass display can show a circle with a handwriting animation of the user. Since artworks that use grass materials for a canvas take time to create, it is difficult for artists to create grass art smoothly. This application allows artists to draw an image on the grass material freely as they would on the tablet.

\subsubsection{Planter Space for Flowers}

Users can determine the color of \textit{ProgrammableGrass} according to the environment and situation. In addition, the resolution can also be changed by adjusting the number of grass modules. Therefore, we created a planter space for flowers using \textit{ProgrammableGrass} as shown in Figure \ref{application}(e). We used two grass modules to create the planter space. In the application, when flowers are put in the planter space, the planter space gets green around the flowers. The users can also change the surface color of the planter space in detail based on the 8-bit levels via the tablet computer, according to the environment and situation.

\section{LIMITATION \& FUTURE WORK}

\subsection{Grass Module}

In the design of the grass pixel,  the subtle differences in the appearance of the grass pixels are due to manually inserting the green grass among the yellow grass. If this process could be mechanized, the appearance of the grass pixels could be aligned for a more diverse perspective. In addition,  the slit structure was adopted to insert the green grass into the gaps of the yellow grass easily. However, the appearance of the grass pixel varies depending on the viewing angle. We suppose that a yellow grass component with holes in a matrix-like grid will reduce differences in appearance depending on the viewing angles. 

The enclosure of the grass module was developed to expand the resolution's width using multiple grass modules. If the enclosure is designed to increase the resolution's width and height, the users can adjust the resolution of \textit{ProgrammableGrass} freely. In addition, \textit{ProgrammableGrass} was assembled using up to four grass modules. As the number of grass modules increases, signal issues such as delay and noise may arise. Therefore, when more grass modules are used, it is necessary to design  electronic circuits considering  the signal issues according to the usage environment.  For example, this solution includes ferrite beads as noise filters, and employing metal shielding. 

The gesture-based animation described in Section \ref{demo} shows a slight delay in the display of \textit{ProgrammableGrass} relative to the gesture input. This delay is due to the Bluetooth communication speed between \textit{ProgrammableGrass} and the tablet computer. It can be resolved by using a wired connection or a more powerful wireless communication method, such as Wi-Fi.

Grass fields are usually used as playing fields and elements that improve green landscapes. We believe that \textit{ProgrammableGrass} will become familiar with social environments, and the relationship between \textit{ProgrammableGrass} and users will become closer when \textit{ProgrammableGrass} is sturdy enough to be stepped on by the users and small enough to be easily installed. Standing on the current artificial grass display is possible due to the aluminum frame. However, the 3D-printed yellow grass blades might break  due to the Polylactic acid (PLA) materials. Therefore, the durability of the  yellow grass  can be further enhanced using resilient artificial grass materials like polyethylene  and thermoplastic polyurethane (TPU).  It is also needed to design how to control the grass length when people are on the artificial grass display. 

Furthermore, to expand the artificial grass display to the size of a sports field and a park, it is important to improve the resolution and display size of a pin display, which is a fundamental technology of \textit{ProgrammableGrass}. Additionally, if \textit{ProgrammableGrass} were to be considered as part of the infrastructure for a sustainable society in the future, energy efficiency would also become necessary. The extensive use of actuators in a pin display leads to high power consumption and increases costs. Therefore, by focusing on minimizing development costs and improving energy efficiency, we can not only enhance the resolution and display size of \textit{ProgrammableGrass} but also contribute significantly to the realization of a sustainable society. For example, MagneShape \cite{MagneSahpe} is an example of technology where a pin display can be operated without electricity and at a lower cost by moving a magnetic sheet against pins embedded with magnets.

\subsection{Grass Color Calibration System}

The green grass lengths can correspond to the 8-bit levels using the grass color calibration system.  It is also possible to display 16-bit images on \textit{ProgrammableGrass} by creating a correspondence table that focuses on 16-bit levels instead of 8-bit levels.  However, when the difference between the grass lengths of adjacent 8-bit  or 16-bit  levels is less than 10/73 [mm], those grass lengths are converted to the same counts using a gain value of 7.3. This problem arises due to the resolution of the grass module's magnetic rotary encoder; utilizing a high-resolution rotary encoder would solve this issue.

In this paper, we conducted experiments on the 8-bit calibration of a single grass pixel under various lighting conditions including standard illumination defined by ISO, the classroom, the meeting room, and the dimly lit space. However, the calibration experiments conducted using the tablet PC in this study are subject to limitations, as they require redoing the experiments in the actual field every time the placement of the grass display is changed. To achieve a more straightforward and flexible 8-bit calibration, we are focusing on a method that completes the calibration in a virtual space. This approach allows for calibration in diverse environments by simply changing the High Dynamic Range Image (HDRI) for the lighting environment. Indeed, we have previously proposed a simple calibration method for grass pixels using a virtual space \cite{mizuno_2023}. In future work, we are planning to develop a system that can realize an 8-bit grass color calibration in a virtual space.

In addition to this, it is important to test the effectiveness of the calibration system in actual  outdoor environments.  Unlike indoor environments, outdoor settings are characterized by very strong illumination, as well as constantly changing illumination and white balance. These changes are due to the movement of the sun and shifts in weather conditions. The current 8-bit calibration system is based on the assumption of constant illumination and white balance. Therefore, addressing fluctuating illumination and white balance is essential for realizing an 8-bit grass color calibration in outdoor environments in the future. 

\subsection{Application \& Evaluation}

In Subsection \ref{SectionApp}, we introduced the interactions with \textit{ProgrammableGrass} via the tablet PC. Along with these functionalities, if the resolution and display size of \textit{ProgrammableGrass} are improved, it would enable the creation of interactive art such as calligraphy art on the grass. Moreover, if \textit{ProgrammableGrass} is installed as part of the real-world infrastructure, urban landscape designers would be able to flexibly determine the color of the grass to match the surrounding buildings, the ambiance of the city, and the seasons. 

In real-life situations, people engage with grass, such as having picnics in a park with family and friends, and this interaction with grass facilitates smooth communication among people. To enhance the relationship between users and grass, it is important to have functionalities that allow interactions such as drawing illustrations or controlling animations directly by touching the grass, without a tablet PC.

Modern artificial grass materials are available in various colors, such as red and blue. Such artificial grass materials can be used to develop \textit{ProgrammableGrass} with multiple grass colors according to the environment and situation. Furthermore, \textit{ProgrammableGrass} controls the grass color based on spatial additive mixing. Since spatial additive mixing is a phenomenon based on the perceptual model of the human eyes, user studies are important to control the grass color focusing on psychophysical aspects.  In addition, it is also important to confirm whether \textit{ProgrammableGrass} is friendly to green landscapes through user studies.  For example, user studies can be conducted using the Likert scale \cite{likert}, a method well-suited for measuring user emotions, opinions, and evaluations, to determine whether \textit{ProgrammableGrass} can be harmonious with green landscapes.  

\section{CONCLUSION}

We proposed \textit{ProgrammableGrass}, a shape-changing grass display with resolution scalability to swiftly control a grass color based on 8-bit levels. The grass color was changed pixel by pixel from yellow to green using fixed-length yellow grass and adjustable-length green grass.  This novel grass display can linearly control the color of the grass according to 8-bit levels, allowing it to display 8-bit images and videos similar to traditional LCDs. 

We designed and developed a grass module to easily adjust the resolution of \textit{ProgrammableGrass} by combining several grass modules. A PD control with a DC motor operated each grass pixel's length in the grass module. We conducted experiments to evaluate the response speed of controlling the grass length in the grass pixel. The resulting response speed of the grass pixel was up to 1/10 [sec], fast enough to satisfy an animation speed of a limited animation method. 

Since the relationship between the grass color and the grass length is nonlinear, we developed a grass color calibration system using image processing to control the grass color linearly based on 8-bit levels. In the grass color calibration system, a correspondence table between the grass length and the 8-bit level was calculated based on the characteristics between the grass length and color. We conducted evaluation experiments to test whether \textit{ProgrammableGrass} can control the grass color based on the 8-bit levels using the calculated correspondence table. As a result, we found that the calibrated grass pixel can show the grass color linearly at the 8-bit levels from several viewpoints  in several indoor environments including an ISO standard environment, a classroom, a meeting room, and a dimly lit space. In addition, we revealed that the calibration system also reduces pixel-to-pixel color variation by taking into account the characteristics of multiple grass pixels through the experiments.

Finally, we showed several animation  examples on the calibrated grass display using 8-bit grayscale MP4 files. Moreover, we demonstrated several application examples to show the potential of \textit{ProgrammableGrass}.  In our future study, we will improve the appearance and the enclosure size to make \textit{ProgrammableGrass} easier to integrate into social environments and green landscapes.

\section*{Abbreviations}

The following abbreviations are frequently used in this paper:

\begin{table}[h]

  \begin{tabular}{ll}
  LCD        & Liquid Crystal Display                   \\
  DC motor   & Direct Current motor                     \\
  PD control & Proportional-Derivative control          \\
  CIE        & International Commision on Illumination \\
             & (Commission Internationale de l'Éclairage) \\
  CIELAB     & $L^*a^*b^*$ color space defined by CIE   \\
  SPI        & Serial Peripheral Interface              \\
  PCB        & Printed Circuit Board                    \\
  SP         & Desired Setpoint                         \\
  PV         & Process Variable                         \\
  PWM        & Pulse Width Modulation                   \\
  OGCD       & Origin Grass Color Difference            \\
  ISO        & International Standard Organization     
  \end{tabular}
\end{table}

\section*{Acknowledgment}
Parts of this paper include sentences generated by OpenAI's ChatGPT \cite{openaiIntroducingChatGPT}, which were then reviewed and modified by the authors.

\bibliographystyle{unsrt}
\bibliography{access}

\vspace{200pt}

\begin{IEEEbiography}[{\includegraphics[width=1in,height=1.25in,clip,keepaspectratio]{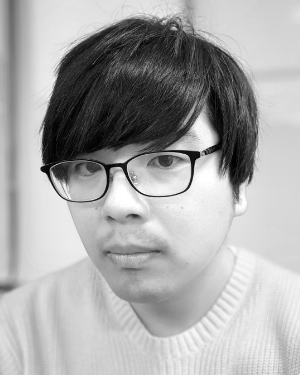}}]{Kojiro Tanaka} received the B.S. degree in Media Sciences and Engineering and the M.S. degree in Informatics from University of Tsukuba, Japan, in 2020 and 2022, respectively. Currently, he is a Ph.D. student at the same university, specializing in Informatics. His research interests include shape-changing display, human computer interaction, and color science.
  \end{IEEEbiography}

  \vspace{-20pt}

  \begin{IEEEbiography}[{\includegraphics[width=1in,height=1.25in,clip,keepaspectratio]{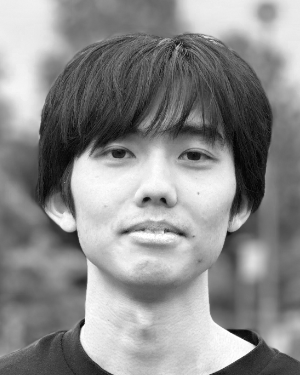}}]{Akito Mizuno} received the B.S. degree in Media Sciences and Engineering, from University of Tsukuba, Japan, in 2022. Currently, he is a master's student at the same university, specializing in Informatics. His current research interests include computer graphics, color science, and robotics.
  \end{IEEEbiography}

  \vspace{-20pt}

  \begin{IEEEbiography}[{\includegraphics[width=1in,height=1.25in,clip,keepaspectratio]{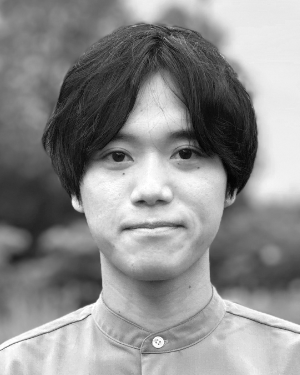}}]{Toranosuke Kato} received the B.S. degree in Media Sciences and Engineering, from University of Tsukuba, Japan, in 2022. Currently, he is a master's student at the same university, specializing in Informatics. His current research interests include human robot interaction, acoustical engineering, and robotics.
  \end{IEEEbiography}

  \vspace{-20pt}

  \begin{IEEEbiography}[{\includegraphics[width=1in,height=1.25in,clip,keepaspectratio]{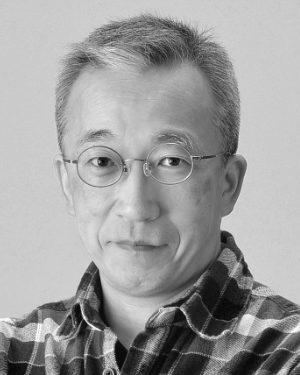}}]{Masahiko Miakwa} is currently a Professor in the Institute of Library, Information and Media Science, University of Tsukuba, Japan. He received B.Eng., M.Eng., and Ph.D. degrees from Osaka University in 1992, 1994 and 2001 respectively. His research interests include robotics, human-Computer interaction, and computer vision. He worked for NTT Access Network Systems Laboratories from 1994 to 2001, NTT Service Integration Laboratories from 2001 to 2002 and was a Lecturer in the Graduate School of Library, Information and Media Studies, University of Tsukuba from 2003 to 2006 and was an Associate Professor from 2006 to 2023. He is a member of RSJ, SICE, SOFT and IEEE.
  \end{IEEEbiography}

  \vspace{-20pt}

  \begin{IEEEbiography}[{\includegraphics[width=1in,height=1.25in,clip,keepaspectratio]{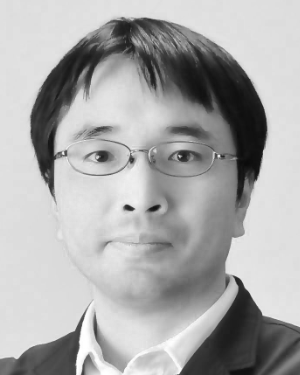}}]{Makoto Fujisawa}  is currently an Associate Professor in the Institute of Library, Information and Media Science, University of Tsukuba since 2021. He received B.Eng., M.Eng., and Ph.D. degrees in mechanical engineering from Shizuoka University in 2003, 2005, and 2008 respectively. He worked for Nara Institute of Science and Technology from 2008 to 2010 and University of Tsukuba from 2011 to 2020 as an assistant professor. His research interests include computer graphics and physics simulation. He is a member of ACM, IEEE
    CS, IIEEJ, IPSJ and VRSJ.
  \end{IEEEbiography}

\EOD

\end{document}